\documentclass[12pt,preprint]{aastex}

\slugcomment{ApJ accepted November 19, 2007}

\shorttitle{Mid-IR spectra of submm galaxies}
\shortauthors{Pope et al.}

\begin{document}

\title{Mid-Infrared Spectral Diagnosis of Submillimeter Galaxies}

\author{Alexandra Pope\altaffilmark{1},
	Ranga-Ram Chary\altaffilmark{2},
	David M.~Alexander\altaffilmark{3},
	Lee Armus\altaffilmark{2},   
	Mark Dickinson\altaffilmark{4},
	David Elbaz\altaffilmark{5},
	David Frayer\altaffilmark{2},
	Douglas Scott\altaffilmark{1},
	Harry Teplitz\altaffilmark{2}
}

\altaffiltext{1}{Department of Physics \& Astronomy, University of British Columbia, Vancouver, BC, V6T 1Z1, Canada}
\altaffiltext{2}{{\em Spitzer} Science Center, MS 220-6 Caltech, Pasadena, CA 91125}
\altaffiltext{3}{Department of Physics, Durham University, Durham, DH1 3LE, UK}
\altaffiltext{4}{National Optical Astronomy Observatory, 950 N. Cherry Ave., Tucson, AZ, 85719} 
\altaffiltext{5}{Laboratoire AIM, CEA/DSM - CNRS - Universit\'e Paris Diderot, DAPNIA/Service d'Astrophysique, CEA Saclay, 91191 Gif-sur-Yvette Cedex, France}

\begin{abstract}
We present deep mid-infrared spectroscopy with the {\it Spitzer Space Telescope} of 13 submillimeter galaxies (SMGs) in the Great Observatories Origins Deep Survey North (GOODS-N) field. We find strong polycyclic aromatic hydrocarbon (PAH) emission in all of our targets which allows us to measure mid-IR spectroscopic redshifts and place constraints on the contribution from star formation and active galactic nuclei (AGN) activity to the mid-IR emission. 
In the high signal-to-noise ratio composite spectrum of SMGs, we find that the hot dust continuum 
from an AGN contributes at most 30\% of the mid-IR luminosity. 
Individually, only $2/13$ SMGs have continuum emission dominating the mid-infrared luminosity; one of these SMGs, C1, remains undetected in the deep X-ray images but shows a steeply rising continuum in the mid-IR indicative of a Compton-thick AGN. 
We find that the mid-IR properties of SMGs are distinct from those of 24$\mu$m-selected ULIRGs at $z\sim2$; the former are predominantly dominated by star formation while the latter are a more heterogeneous sample with many showing significant AGN activity.
We fit the IRS spectrum and the mid-IR to radio photometry of SMGs with template
spectral energy distributions to determine the best estimate of the total infrared luminosity from star formation.
While many SMGs contain an AGN as evinced by their X-ray properties, our multi-wavelength analysis shows that the total infrared luminosity, $L_{IR}$, in SMGs, is dominated by star formation and not AGN activity. 
We find that high redshift SMGs lie on the relation between $L_{IR}$ and $L_{\rm{PAH,}6.2}$ (or $L_{\rm{PAH,}7.7}$ or $L_{\rm{PAH,}11.3}$) that has been established for local starburst galaxies. This suggests that PAH luminosity can be used as a proxy for the star formation rate in SMGs. 
SMGs are consistent with being a short lived cool phase in a massive merger. Although there appears to be an AGN present in many SMGs, it does not appear to have become strong enough to heat the dust and dominate the mid- or far-infrared emission.
\end{abstract}


\section{Introduction}
\label{sec:intro}
The {\it COsmic Background Explorer} ({\it COBE}) revealed that the extragalactic background light at infrared (IR) wavelengths was comparable to that at optical wavelengths indicating that half of the star formation and active galactic nuclei (AGN) emission, averaged over cosmic time, is obscured by dust (e.g.~Puget et al.~1996; Hauser \& Dwek 2001). In the local Universe, the volume density of dusty, infrared luminous galaxies is insufficient to account for this background. Therefore the density of such galaxies must evolve strongly with redshift (e.g.~see Lagache, Puget, \& Dole 2005 for a review). In fact, imaging observations with the {\it Infrared Space Observatory} ({\it ISO}) and the {\it Spitzer Space Telescope} (hereafter {\it Spitzer}) have shown that about 70\% of the comoving star formation rate density (SFRD) at $0.5<z<3$ is obscured by dust (Chary \& Elbaz 2001; LeFloc'h et al.~2005; P{\'e}rez-Gonz{\'a}lez et al.~2005). 

In the local Universe, we refer to these dusty galaxies as luminous infrared galaxies (LIRGs, $11\le \rm{log}\,(L_{\rm{IR}}\,[L_{\odot}])<12$) and ultraluminous infrared galaxies (ULIRGs, $\rm{log}\,(L_{\rm{IR}}\,[L_{\odot}]) \ge 12$) where $L_{\rm{IR}}$ is the total IR luminosity from 8--1000$\,\mu$m (see Sanders \& Mirabel 1996 for a review). It is now becoming clear that IR luminous galaxies are a heterogenous population with selection functions at different mid- and far-infrared wavelengths revealing different subsets of this population (Daddi et al.~2005; Lutz et al.~2005b; Yan et al.~2005; Pope et al.~2006).

Perhaps the most enigmatic members of the high redshift ULIRG population are the submillimeter (submm) galaxies (SMGs). These galaxies were initially discovered in the late 1990s with the Submillimetre Common User Bolometer Array (SCUBA, Holland et al.~1999) on the James Clerk Maxwell Telescope (JCMT) and then later with MAx-Planck Millimetre BOlometer (MAMBO) camera on the Institut de Radio Astronomie Millimetrique (IRAM) 30-m telescope (see Blain et al.~2002 for a review). They have proven to be difficult to study, due to the large beam size of submm observations and the difficulty in finding counterparts at other wavelengths (e.g.~Ivison et al.~2002; Pope et al.~2005). Nevertheless, it has been found that they are very massive systems (Swinbank et al.~2004; Borys et al.~2005; Greve et al.~2005; Tacconi et al.~2006) at $z\sim2$ (Chapman et al.~2005; Pope et al.~2006; Aretxaga et al.~2007) which are thought to be connected to the most massive present day galaxies via an evolutionary sequence (Lilly et al.~1999). A crucial aspect that needs to be
explored in SMGs is what powers their extreme luminosities: AGN activity or star formation?

The mid-IR spectral regime is a particularly good probe of what is producing the total infrared luminosity since it has contributions from both star-formation and AGN each with a unique signature at these wavelengths. The 5--15$\,\mu$m mid-IR spectral energy distribution (SED) can be decomposed into 3 main components (e.g.~Sajina et al.~2007): 
(1) emission features from what are thought to be polycyclic aromatic hydrocarbon (PAH) molecules (Puget \& Leger 1989; Allamandola et al.~1999); (2) power-law (or warm blackbody) emission; and (3) extinction characterized by prominent silicate absorption features (e.g.~Draine 2003). 
The first is understood to be powered entirely by star formation while the second is predominantly emission from the AGN (Genzel et al.~1998). The second component can also contain contributions from hot dust (very small grains) present in the most energetic HII regions (e.g.~Tran et al.~2001). The third component can be found in both starbursts and AGN since it just requires a dust screen around hot dust emission.

Measuring the contribution from ULIRGs to the global star formation rate requires dissecting their IR emission into contributions from starbursts and AGN.
Mid-IR spectroscopy with {\it ISO} made it possible to decompose the various contributors in local ULIRGs with the result that they are predominantly starburst powered, although they also show signs of AGN activity (e.g.~Genzel et al.~1998; Lutz et al.~1999; Rigopoulou et al.~1999; Sturm et al.~2000; Tran et al.~2001). 
In particular, {\it ISO} studies of local ULIRGs showed that at low luminosities, the ULIRGs are powered mainly by a starburst but at the very highest luminosities ($L_{\rm{IR}}\gtrsim3\times10^{12}\rm{L}_{\odot}$), the dominant emission source is the AGN (Lutz et al.~1998; Tran et al.~2001). From this one might expect that SMGs, with their extreme IR luminosities ($L_{\rm{IR}}\sim6\times10^{12}\rm{L}_{\odot}$, Pope et al.~2006), are primarily AGN-dominated.  

The {\it Spitzer} Infrared Spectrograph (IRS; Houck et al.~2004), with its sensitivity and spectral resolution, can extend this effort outside the local Universe, allowing much larger samples to be observed to much fainter flux limits. The IRS can detect emission features and continuum in ULIRGs out to $z\sim4$ (Valiante et al.~2007). The 3 mid-IR spectral components listed above have been fit to the {\it Spitzer} IRS spectra of ULIRGs at various redshifts to determine the AGN contribution to the mid-IR luminosity (e.g.~Sajina et al.~2007). Although X-ray imaging is very effective at identifying the presence of an AGN in a galaxy, it is subject to  obscuration in extreme, Compton-thick sources. The X-ray luminosity provides a direct measurement of the power of the AGN at X-ray energies but it is necessary to make assumptions about the geometry of the obscuring dust to
determine the bolometric luminosity of the AGN (e.g.~Alexander et al.~2005). Furthermore, AGN indicators at UV and optical wavelengths can be very obscured in dusty SMGs (Swinbank et al.~2004; Chapman et al.~2005). Mid-IR spectroscopy, by directly detecting the thermally re-radiated emission from the obscuring dust, provides one of the most effective ways to identify AGN, including Compton-thick sources, and measure their contribution to the infrared luminosity (e.g.~Lutz et al.~2004). 

Two previous studies have presented IRS spectroscopy of SMGs. Menendez-Delmestre et al.~(2007) presented IRS spectra of 5 radio-detected, spectroscopically confirmed SMGs. They found that four SMGs with lower redshifts ($z<1.5$) showed PAH emission and a composite spectrum of these four sources fits well to a scaled M82 spectrum plus a power-law component. They conclude that these systems host both star formation and AGN activity. Their fifth source at $z=2.4$ was found to have a more AGN-type spectrum. Valiante et al.~(2007) discussed IRS spectra of 13 bright SMGs, 9 of which are detected with the IRS (the first two spectra from this sample were presented in Lutz et al.~2005a). This sample was more representative of the redshift distribution of SMGs. They also found that SMGs are likely to be mainly starburst powered. 
Sajina et al.~(2007) used IRS spectra to study a sample of {\it Spitzer} 24$\,\mu$m-selected galaxies at $z\sim2$. This sample was selected to have $S_{24\,\mu m}>900\,\mu$Jy and very red 24-to-8$\,\mu$m and 24-to-$R$ colors (Yan et al.~2007).
 75\% of this sample consists of continuum-dominated sources, with some weak PAH emission indicative of star formation activity. They conclude that these sources display mid-IR spectra similar to those of the most luminous local ULIRGs. Follow-up studies of these galaxies have shown that they are not generally (sub)mm detected sources ($S_{1.2\rm{mm}}\lesssim 1.5\,$mJy for $33/40$ sources, Lutz et al.~2005b).

In this paper we present {\it Spitzer} IRS spectra of a sample of SMGs in GOODS-N which have extensive multi-wavelength information from X-ray to radio wavelengths. This allows us to produce full spectral energy distributions and to calculate the bolometric luminosity. Our {\it Spitzer} IRS observations go substantially deeper ($S_{24\,\mu m}>200\,\mu Jy$) than previous work. This enables a study of
the typical blank-field submm galaxy population which have $S_{850\,\mu m}>2\,$mJy. The main goal of our program is to decompose the mid-IR spectra into star formation and AGN components, in order to infer how much of the total infrared luminosity is coming from each. From the mid-IR spectroscopy we are also able to obtain spectroscopic redshifts and measure the luminosities and equivalent widths of individual PAH lines to investigate various line diagnostics. We compare our measurements with line strengths in other star-forming galaxies to put constraints on the evolution of SMGs and the relationship between star formation and black hole growth. 

This paper is formatted as follows. Our {\it Spitzer} observations are described in Section 2 and the data analysis is explained in Section 3. Section 4 presents the main results of the IRS spectroscopy including redshifts, spectral decomposition and AGN classification. In Section 5 we fit the full SEDs of these SMGs to determine their total IR luminosities. We explore correlations between the PAH and IR luminosities in Section 6. We discuss our results in the context of the role that SMGs play in galaxy evolution in Section 7. Finally, our conclusions are listed in Section 8. We include an appendix which discusses the noise properties of our {\it Spitzer} IRS spectra and includes notes on individual sources. 

All magnitudes in this paper use the AB system. We use a standard cosmology with $H_{0}=73\,\rm{km}\,\rm{s}^{-1}\,\rm{Mpc}^{-1}$, $\Omega_{\rm{M}}=0.3$ and $\Omega_{\Lambda}=0.7$. 

\section{Observations}
\label{sec:obs}
The {\it Spitzer} IRS observations presented here were taken as part of {\it Spitzer} program GO-20456 (PI: R.~Chary). The full sample consists of high redshift SMGs, AGN and optically faint 24$\,\mu$m sources in the Great Observatories Origins Deep Survey (GOODS, Dickinson et al.~2003) North field. 
The GOODS-N field is one of the most extensively studied regions of the sky, with deep data existing across all wavebands including: {\it Chandra} 2$\,$Msec X-ray observations (Alexander et al.~2003); deep {\it HST} optical imaging using the Advanced Camera for Surveys (ACS) in four bands, $B_{\rm{435}}$, $V_{\rm{606}}$, $i_{\rm{775}}$ and $z_{\rm{850}}$ (Giavalisco et al.~2004); and deep {\it Spitzer} imaging at five infrared wavelengths (Dickinson et al.,~in preparation). More recently, GOODS-N has been surveyed with IRS peak-up observations at 16$\,\mu$m (Teplitz et al.~2005) and MIPS imaging at 70$\,\mu$m (Frayer et al.~2006).
In addition to the extensive space-based imaging campaign, many ground-based programs have also targeted this field for imaging and spectroscopy making it one of the best field for investigating the role of AGN and star formation in high redshift galaxies. 
Here we present the IRS spectroscopy for the SMGs from this project. Results from the other sources will be presented in other papers.

Our sample of SMGs was primarily selected from sources detected in the GOODS-N SCUBA `super-map' 
with secure multi-wavelength counterparts (Borys et al.~2003; Pope et al.~2006; Pope 2007).
We imposed a 24$\,\mu$m flux cut of $S_{24}>200\,\mu$Jy in 
order to ensure that we achieve adequate signal to noise (SNR) on the continuum in the IRS
spectra. Roughly half of the secure counterparts in Pope et al.~(2006) meet this 24$\,\mu$m flux cut which yields
a sample of 10 IRS targets (after removing sources which are already observed with the IRS in other programs). To expand our sample slightly, we also included three SCUBA sources from Chapman et al.~(2005). These three sources are not detected at $>3.5\sigma$ in the `super-map' but they are detected in SCUBA photometry observations with a lower statistical significance. 

\begin{figure*}
\epsscale{.80}
\plotone{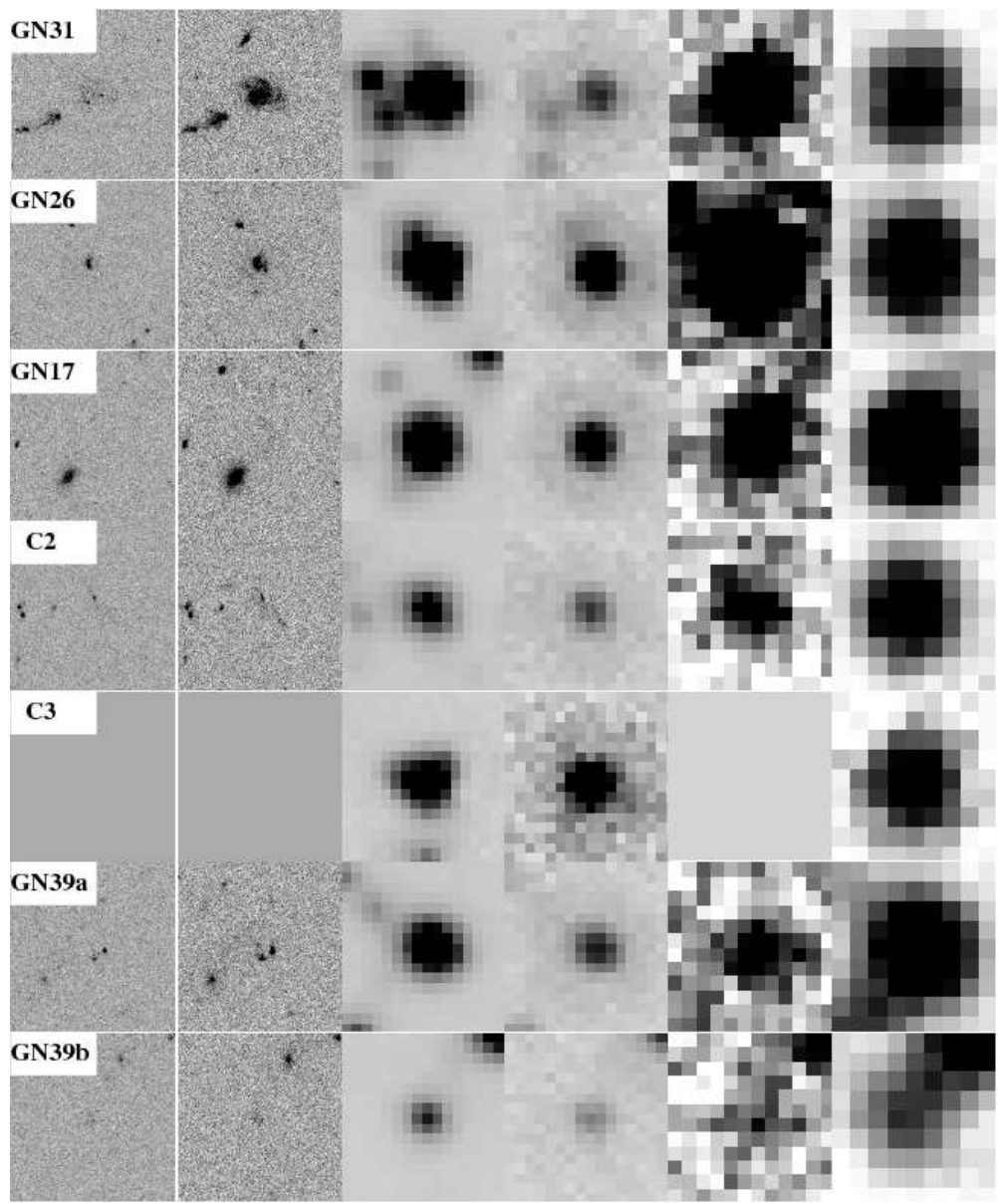}
\caption{Images of SMG targets, in order of increasing redshift. Images are $10^{\prime\prime}\times10^{\prime\prime}$ (roughly the size of the aperture of the IRS observations) at $B$, $i$, 3.6, 8.0, 16 and 24$\,\mu$m from left to right. 
Data are not available in some of these wavebands for source C3 since it is outside the {\it HST} GOODS-N coverage. 
GN04, GN07, GN19 and GN39a and b, have two sources within the slit and which contribute flux in the IRS spectra (See text for details).
For the other sources, the 8, 16 and 24$\,\mu$m images show that only the primary targets are bright in the mid-IR.
\label{fig:post}}
\end{figure*}

Our 13 SMG IRS targets have 850$\,\mu$m and 24$\,\mu$m flux densities of 2--10$\,$mJy and 200--1200$\,\mu$Jy, respectively. This sample is expected to be representative of the submm population discovered in existing blank-field 850$\,\mu$m imaging. There is no evidence supporting the idea that SMGs which are fainter than 200$\,\mu$Jy at 24$\,\mu$m are fundamentally different besides lying at slightly higher redshifts (e.g.~Pope et al.~2006). We list the positions and mid-IR fluxes for these 13 SMGs in Table \ref{tab:targets}. SMG IDs starting with 'GN' and 'C' are from the SCUBA super-map (Pope et al.~2006; Pope 2007) and Chapman et al.~(2005), respectively, and the 850$\,\mu$m fluxes can be found in these papers. The majority of the SMGs are undetected in the deep {\it Spitzer} $70\,\mu$m images (see also Huynh et al.~2007). 11/13 SMGs have optical spectroscopic redshifts while the other two have photometric redshifts. The redshift distribution of our sample is consistent with that of all SMGs (Chapman et al.~2005; Pope et al.~2006). Fig.~\ref{fig:post} shows multi-wavelength 
images, which span roughly the width of the IRS slit, of our targets. In some cases there are 
several optical or IRAC sources in the aperture.
For GN04, GN07 and GN19, the two IRAC sources are both expected to be associated with the submm emission since both components are at the same redshift (see Pope et al.~2006). The IRS spectra of these 3 SMGs will have contributions from both components.  
For the remaining sources, the 8, 16 and 24$\,\mu$m images show that only our primary targets are producing the mid-IR emission which we will detect with the IRS.
\clearpage
\clearpage
{\plotone{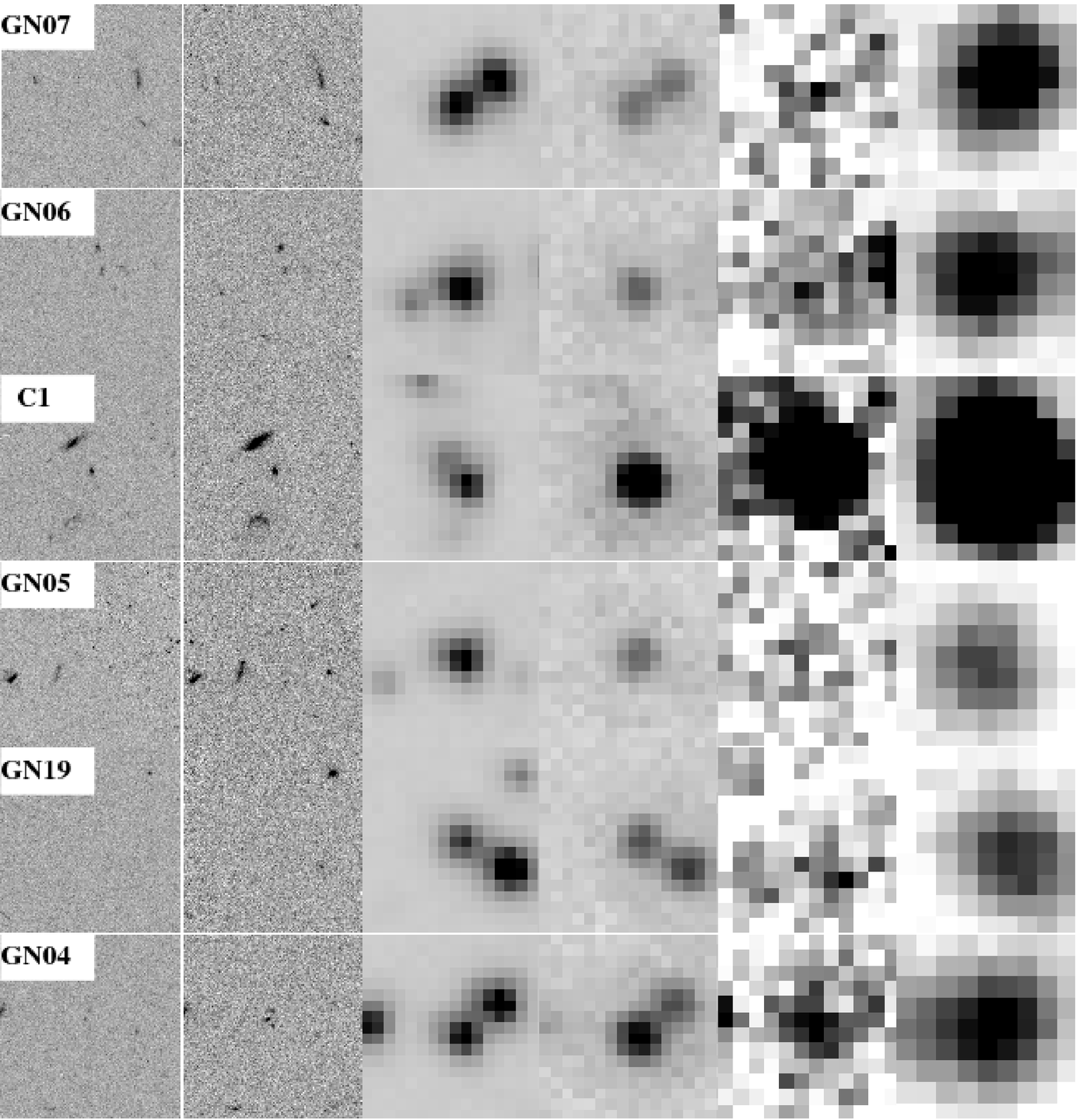}}\\
\centerline{Fig. 1. --- Continued.}
\clearpage

The {\it Spitzer} IRS observations are taken in
spectral staring mode which observed the target at two nod positions within the slit. 
We observed in low resolution ($R=\lambda/\Delta\lambda\sim$100) using the Short-Low 1 (SL1; 7.4--14.5$\,\mu$m), Long-Low 2 (LL2; 14.0--21.3$\,\mu$m) and/or Long Low 1 (LL1; 19.5--38.0$\,\mu$m) orders, depending on the redshift of the source. We chose the wavelength coverage in order to include one or more of the main PAH features at 6.2, 7.7 and 11.3$\,\mu$m. Integration times varied depending on the flux of the target at 24$\,\mu$m.
We used blue peak-up acquisition on nearby, isolated, bright 2MASS stars.
Our observations were taken in April and May 2006.

Table \ref{tab:obs} shows our list of targets and integration times.

\section{Data analysis}
\label{sec:analysis}
We started with the S14.0.0 $\it Spitzer$ IRS pipeline data\footnote{\tt{http://ssc.spitzer.caltech.edu/irs/dh/}} which produces 2-dimensional (2D) Basic Calibrated Data (BCD) files. Note that our targets are not resolved with {\it Spitzer} at these wavelengths and therefore they can be treated as point sources.

At mid-IR wavelengths, the dominant source of noise is the sky background, mostly from zodiacal light. Since we have long integrations of very faint targets, additional data reduction steps are required to alleviate the detector and sky noise.
The first step is to identify and clean the rogue pixels. 
There are a number of hot pixels in the IRS arrays ($\sim15\%$). Rogue pixel masks were created from the campaign masks and from examining each 2D spectral image individually. 
We used IRSCLEAN\footnote{\tt{http://ssc.spitzer.caltech.edu/archanaly/contributed/irsclean /IRSCLEAN\_MASK.html}} to replace the values of the hot pixels by extrapolating from the surrounding pixels. Note that IRSCLEAN fails when there is a group of several hot pixels together, so we made sure to exclude these from the spectral extraction. The fraction of pixels in the 2D files which remain unusable after IRSCLEAN is only $\sim2\%$.

The next step is to remove the latent charge build-up on the array. In long IRS integrations, it has been found that latent charge builds up with time, despite the detectors being reset at the end of each integration (see Teplitz et al.~2007). This build-up is different depending on the wavelength ($y$ position on the 2D spectral image)\footnote{\tt{http://ssc.Spitzer.caltech.edu/irs/documents/irs\_ultra deep\_memo.pdf}}.
We found charge build-up on the arrays for integrations of more than 1 hour in all LL observations. For the SL1 observations, we only saw the latent charge build-up for a few targets, even though all of these observations were greater than 1 hour. 
The latent charge build-up was fit by a polynomial and the temporal component subtracted off.

We explore several methods for removing the sky background including: a) subtracting the average spectrum at each nod position; b) subtracting individual observations from opposite nod positions and then averaging; c) subtracting a `supersky' created from all files from that Astronomical Observation Request (AOR); and d) subtracting a supersky created from all our data taken in a given campaign. In all methods, we mask out all sources which fall within the IRS slit identified from the deep 24\,$\mu$m imaging of the GOODS-N field. The supersky subtraction was carried out with and without normalizing to the median in each image. 
The third method of subtracting a normalized supersky created from all observations from that AOR gave the lowest residual sky noise (and thus highest SNR) in the final spectrum therefore this is what we used to remove the sky from all our observations. 

After the sky was removed from the individual data files, we then coadded them using a clipped median. The coadds were checked again for persisting hot pixels and cleaned once more using IRSCLEAN. These steps produced two 2D spectra for each target, one for each nod position. 

We used the SPitzer IRS Custom Extraction (SPICE\footnote{\tt{http://ssc.spitzer.caltech.edu/postbcd/spice.html}}) to extract the 1D spectra at each nod position. We used a narrow extraction window of 2 pixels ($\sim$10 arcseconds for LL observations) which was constant as a function of wavelength. We found this produced higher SNR spectra than using an extraction window of increasing width, starting with 2 pixels at the blue end. Using the optimal extraction option in SPICE, we found comparable results to the 2 pixel, fixed width, extraction. For each science spectrum, we also extracted a residual sky spectrum, offset from any sources, to provide a level of the error in our final spectrum as a function of wavelength. The uncertainty is calculated as the standard deviation of 10 pixels surrounding each wavelength in the residual sky spectrum. For each source, the 1D spectra at each nod position were then averaged together to produce the final spectrum for each source. Further details about the final noise in our IRS spectra are given in Appendix~\ref{app:rms}.

\begin{figure}
\epsscale{.80}
\plottwo{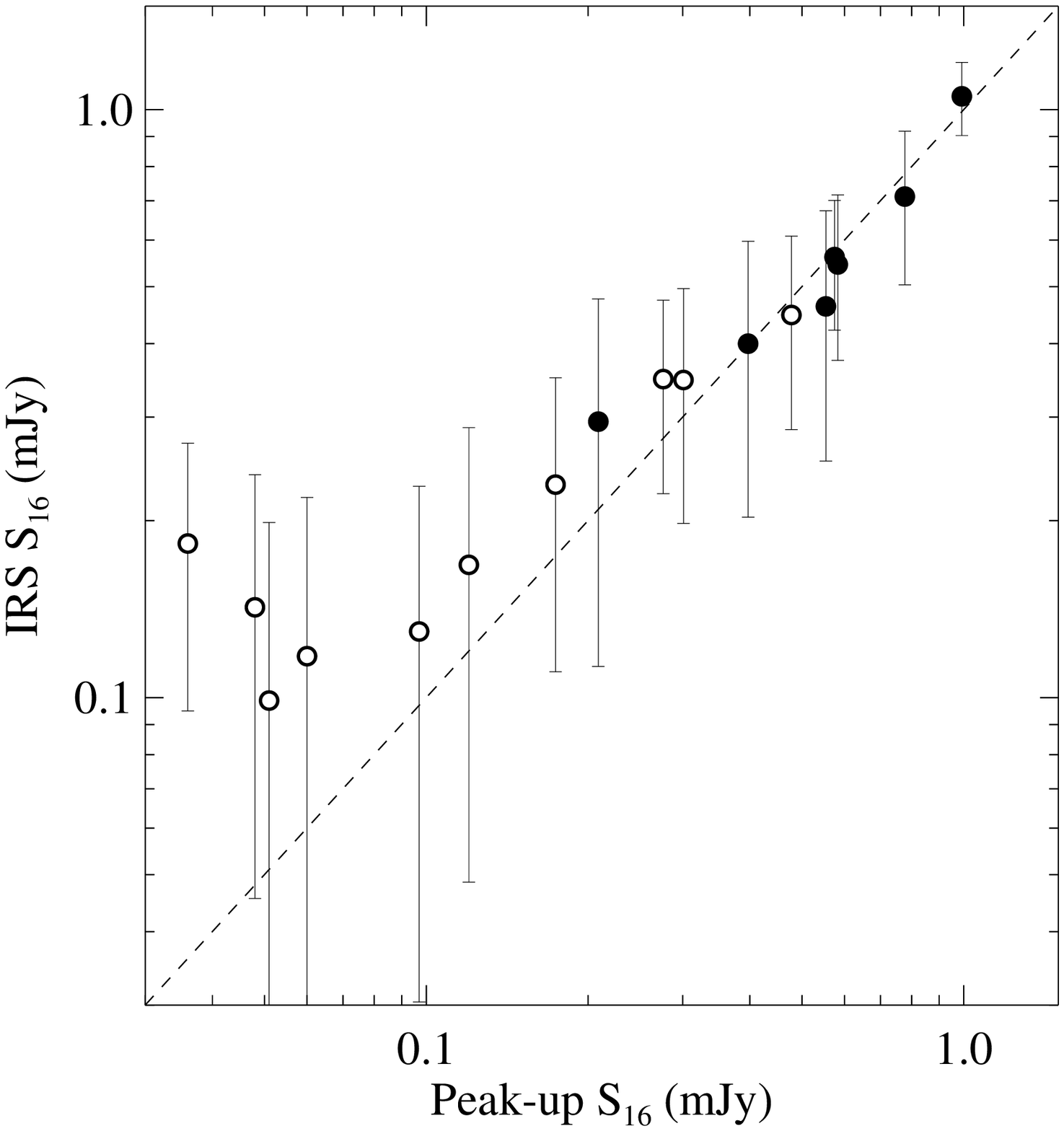}{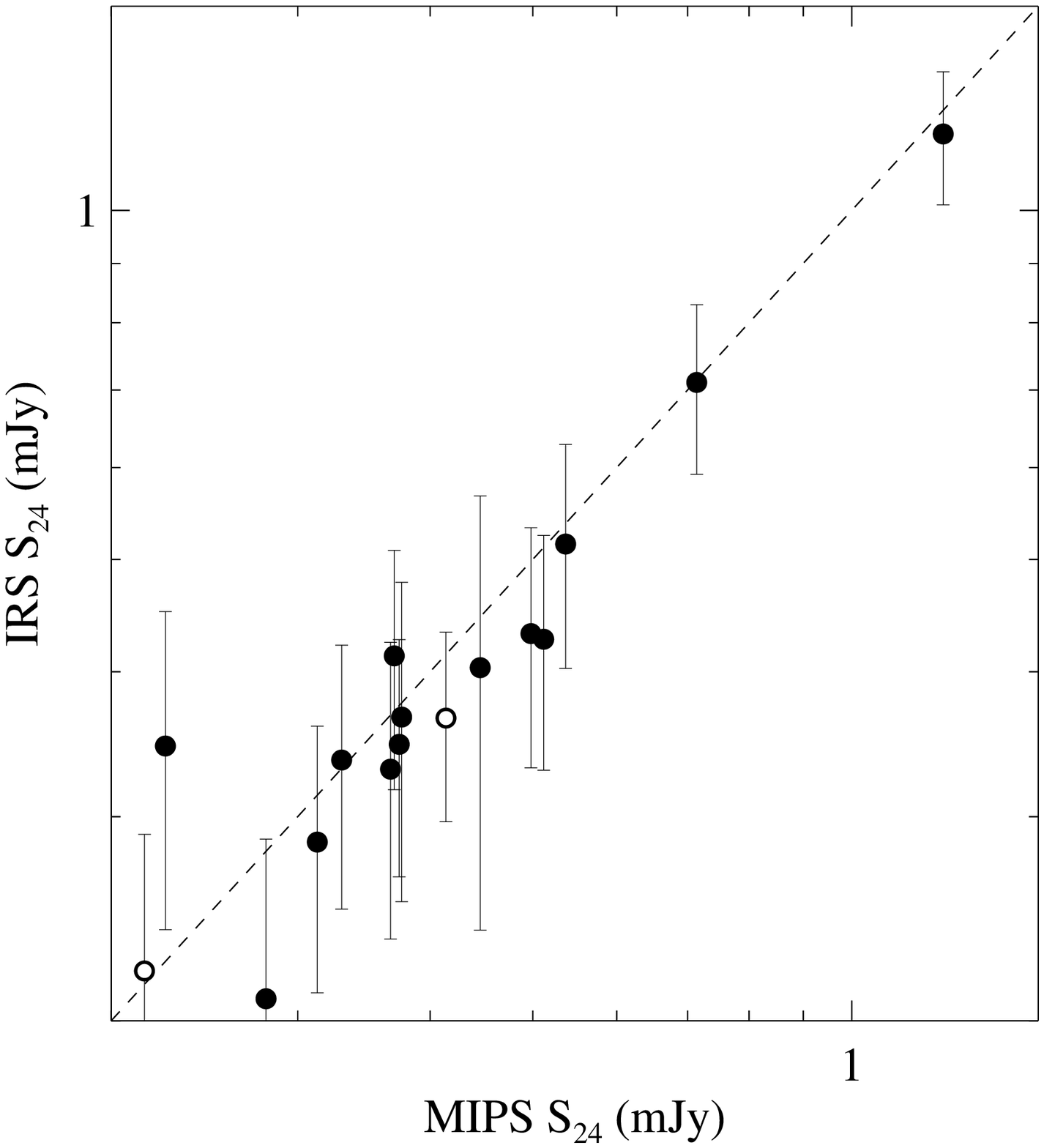}
\caption{Calibration of IRS spectra compared to the MIPS and peak-up photometry. MIPS photometry is from the GOODS $24\,\mu$m imaging (Chary et al.~in preparation) and the 16$\,\mu$m peak-up imaging is from Teplitz et al.~(2005). Most of the 16$\,\mu$m IRS fluxes are systematically higher because roughly half of the 16$\,\mu$m filter falls outside of the LL2 IRS coverage and so the IRS flux is overestimated, since the spectrum actually drops off at lower wavelengths. These sources are indicated as the open circles. For sources with SL1 and LL2 observations, there should be no such discrepancy, which we confirm.
The solid point in the right panel which is significantly above the line is GN39b whose IRS spectrum is known to contain a contribution (1/3 of the flux) from GN39a. 
 \label{fig:cal}}
\end{figure}

In order to calibrate the final IRS spectra, we reduced standard star calibrator observations from the same campaign as our observations, in exactly the same way that we reduced the science observations. We then extracted the spectrum for the calibrator using our 2 pixel width and the default SPICE width in order to determine the calibration factor as a function of wavelength. This calibration was verified against the broad-band photometry at 8, 16 and 24$\,\mu$m and we found very good agreement (see Fig.~\ref{fig:cal}). 

\clearpage
\clearpage

\subsection{Spectral line measurements}
\label{ap:meas}
In order to measure the individual PAH line luminosities we used the IDEA (ISAP\footnote{The ISO Spectral Analysis Package (ISAP) is a joint development by the LWS and SWS Instrument Teams and Data Centers. Contributing institutes are CESR, IAS, IPAC, MPE, RAL and SRON.}) environment in the SMART\footnote{SMART was developed by the IRS Team at Cornell University and is available through the Spitzer Science Center at Caltech.} software package (Higdon et al.~2004). For each of the 4 main PAH lines (6.2, 7.7, 8.6 and 11.3$\,\mu$m), we isolated a region around each line and simultaneously fit the line and continuum. Both the center and the width of the line were allowed to vary. All measured line widths were consistent within the errors with the line widths measured in IRS spectra of local star forming galaxies (Smith et al.~2007). Note that the errors we calculated from the residual sky were propagated in the fitting of individual lines. All SMG line measurements used in this paper including 3$\sigma$ upper limits when the line is undetected in the IRS spectrum are listed in Table \ref{tab:lines}.

The biggest uncertainty in comparing PAH line luminosities from different samples comes from differences in how the spectral lines are measured, more specifically, how the continuum is estimated for each line. This can lead to variations in the line flux and equivalent width of up to a factor of 4 (see Sajina et al.~2007; Smith et al.~2007).
Our method of measuring the line strengths gives similar results to the cubic spline continuum fits performed in Brandl et al.~(2006, see next paragraph).
This method is known to underestimate the flux of the PAH lines (particularly the 7.7$\,\mu$m line) compared to methods which fit all PAH lines simultaneously (e.g.~Sajina et al.~2007; Smith et al.~2007) since the continuum we adopt is higher. Therefore, we took care to measure the PAH lines in any comparison samples in the exact same way as we did for the SMGs. 

In this paper, we compare the SMG PAH line luminosities to those of other galaxy populations. Brandl et al.~(2006) presented IRS spectra of 22 local starburst galaxies with infrared luminosities ranging from $5\times 10^{9}$--$5\times 10^{11}\rm{L_{\odot}}$. These galaxies are known to have little or no AGN contributing to their infrared luminosity. Because the choice for estimating the continuum under the PAH lines makes a significant difference to the results, we have measured the lines for the starburst galaxies in exactly the same way as we did for the SMGs. Another interesting comparison sample is the local ULIRG galaxies. Armus et al.~(2007) presented IRS spectroscopy of 10 ULIRGs from the Bright Galaxy Sample (BGS, $S_{60}>5.4\,$mJy, Soifer et al.~1987) and the IRS spectra for a sample of over 100 ULIRGs was recently presented in Desai et al.~(2007) and Farrah et al.~(2007). The local ULIRGs show a broad range of mid-IR properties and a mix of AGN and starburst mid-IR emission. However, we did not have access to the IRS spectra for the full sample of local ULIRGs in order to measure the line luminosities using our method therefore we could not include them in this paper. 

We measured the PAH lines in 10 of the Brandl et al.~(2006) starburst galaxies to verify that our method is consistent with their measurements. On average the luminosities of the 6.2 and 11.3$\,\mu$m lines we measure are consistent with the values listed in Brandl et al.~(2006), but the 7.7$\,\mu$m line luminosity is 1.5 times larger on average from our measurements. This is probably because we are using a straight line for the continuum as opposed to a spline. We have corrected the measurements of the 7.7$\,\mu$m line luminosity for the Brandl et al.~(2006) starbursts for this factor of 1.5 for all plots in this paper. 
For the Brandl et al.~(2006) starbursts we also applied a correction to the line measurements to account for the fractional flux found within the IRS apertures (see Table 2 of Brandl et al.~2006) when comparing to the unresolved SMGs. The error on the line fluxes for the Brandl et al.~(2006) starbursts is dominated by the calibration error which is on the order of 10\% (less than the size of the symbols in most of the plots in this paper).

\section{Results}
In this section we use the mid-IR spectra to explore the nature of SMGs. We start by using the identification of PAH lines to estimate redshifts. We then proceed to decompose the mid-IR spectra into starburst and AGN components; starting first with a composite SMG spectrum since the SNR is higher and then proceeding to decompose each individual SMG spectrum. We then use the information from the spectral decomposition to classify the fractional contribution of AGN to the mid-IR luminosity and compare this with other AGN indicators such as optical spectroscopy and X-ray emission. 
These results are then used to guide the fitting of the full infrared SED to constrain the total infrared luminosity in the Section 5. 

\begin{figure*}
\epsscale{.80}
\plotone{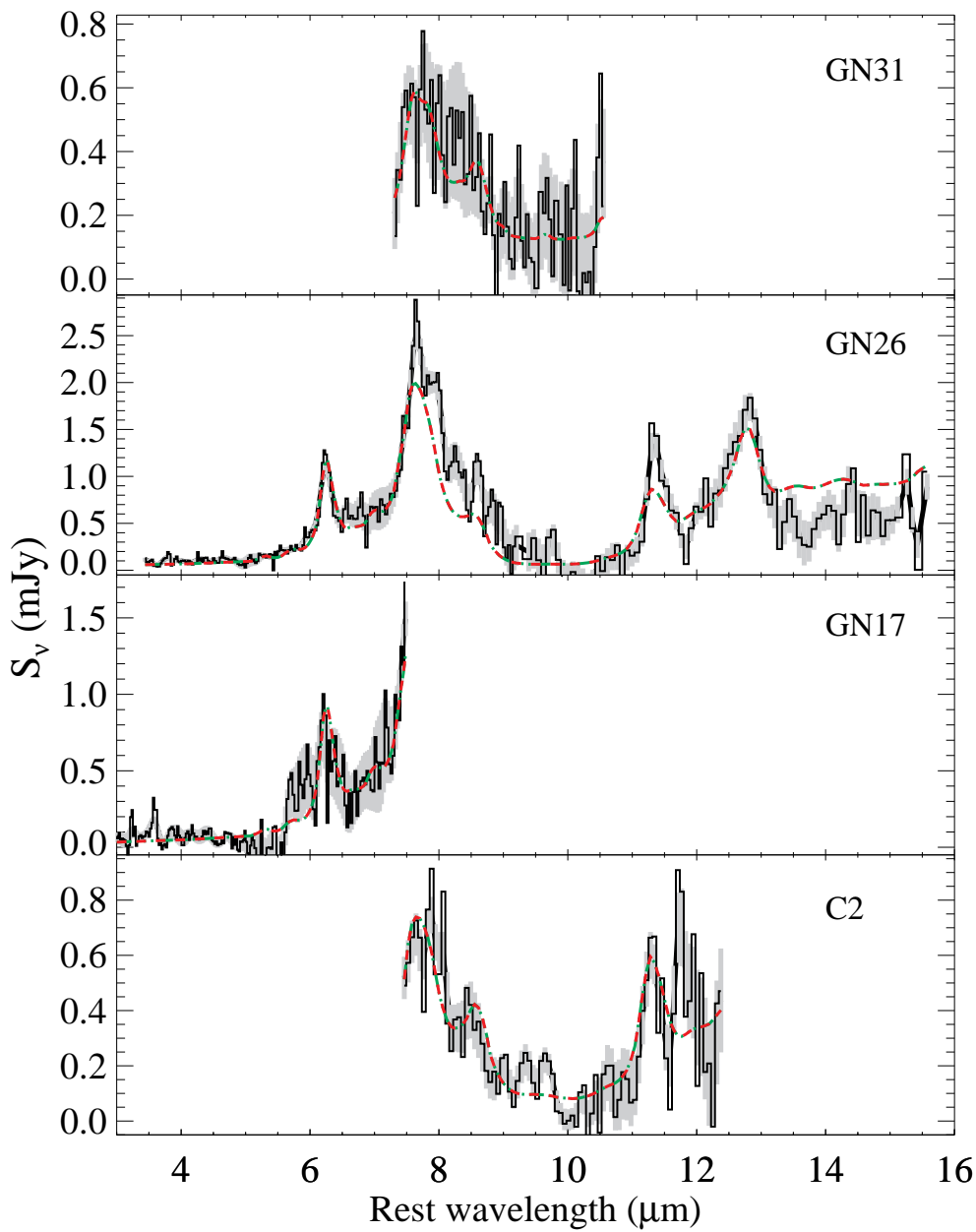}
\vspace{0.5in}
\caption{IRS spectra of SMGs. The solid black curves are the raw (unsmoothed) IRS spectra, while the shaded region is the associated $1\sigma$ noise from the sky background. The red dashed curves show the best-fit SED which is made up of an extincted  power-law component (blue dashed, consistent with zero in most cases) and an extincted starburst component (green dashed). Recall that the extinction curve is not monotonic in wavelength and contains various features, the most notable being the 9.7$\,\mu$m silicate feature.
\label{fig:spec}}
\end{figure*}

Fig.~\ref{fig:spec} shows the rest-frame (unsmoothed) IRS spectra for the 13 SMGs. PAH emission lines are present in all sources. Despite the relatively low SNR in the spectra, all SMGs seem to show very similar mid-IR spectral shapes with the exception of C1. While this SMG clearly shows PAH emission, it is superimposed on a steeply rising continuum. We will discuss this source further in Section \ref{ap:sources}.
\clearpage
\clearpage
{\plotone{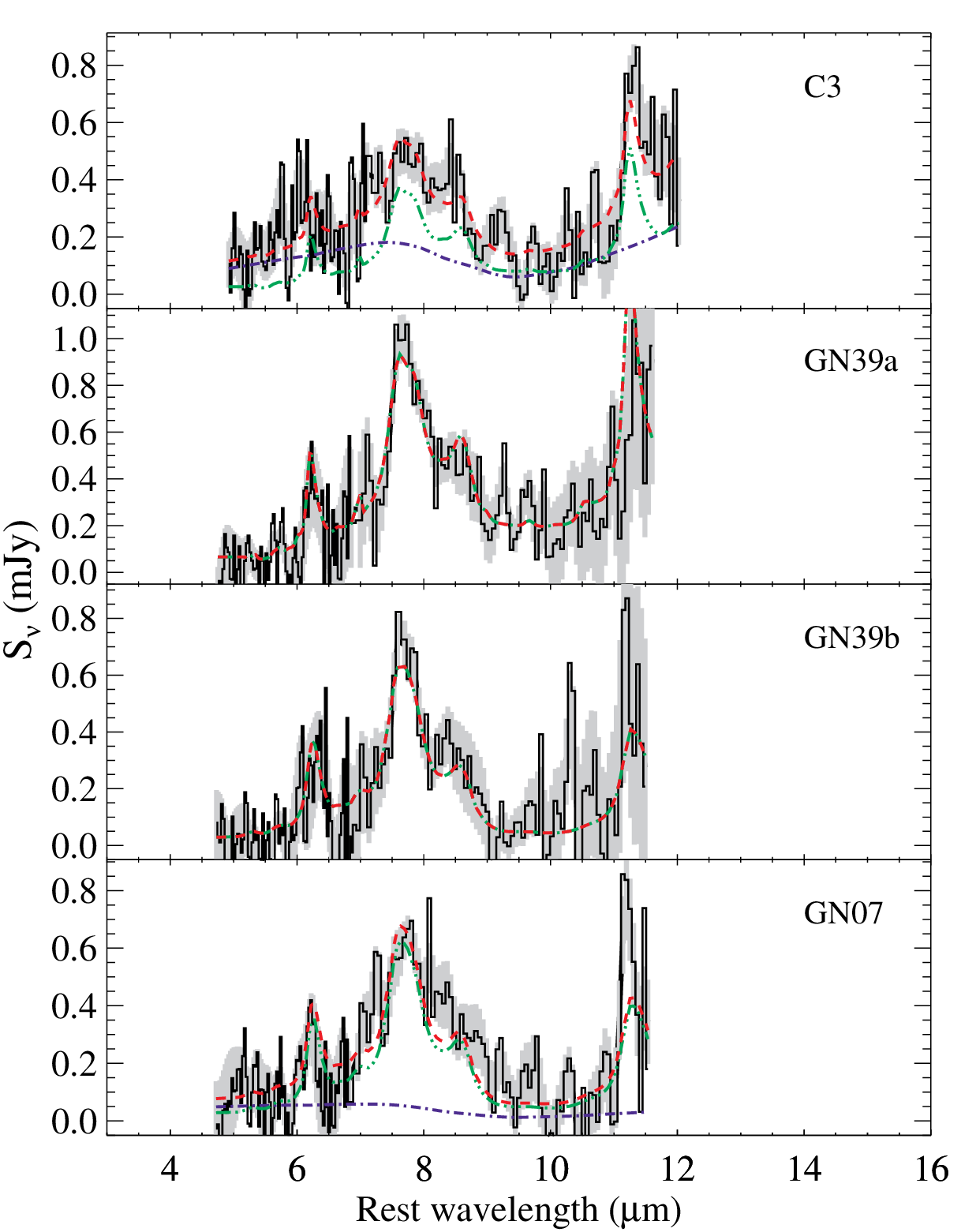}}\\[5mm]
\centerline{Fig. 3. --- Continued.}
\clearpage
\thispagestyle{empty}
\vspace*{-20mm}
{\plotone{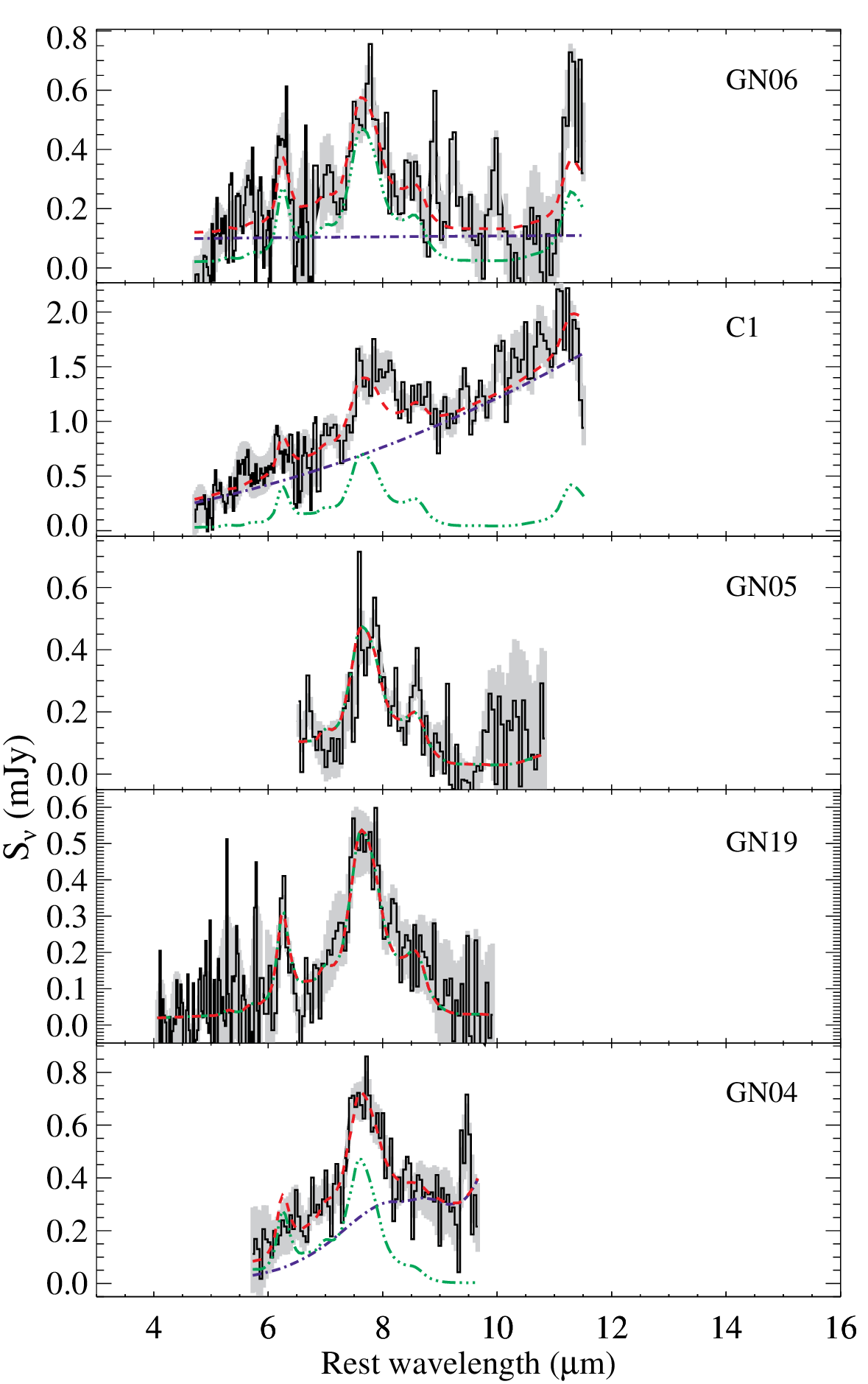}}\\[5mm]
\centerline{Fig. 3. --- Continued.}
\clearpage

\subsection{Redshifts}
Given the presence of PAH features in all of our sources we are able to extract redshifts from the IRS spectra.
In order to calculate the redshift and its associated error from the IRS data we follow the following steps.
We assumed the following rest-frame wavelengths for the main PAH features; 6.22, 7.71, 8.61 and 11.27$\,\mu$m (Draine \& Li 2007). 
For each PAH line, we find the center of the Gaussian fit to the line from SMART and calculate the redshift. The redshift for each source is calculated by averaging the redshift estimates from each PAH line. 
The total error in the redshift is the quadratic sum of the deviation due to the spread in the redshift from the different PAH lines and the centroid uncertainty. For sources whose spectra only contain 1 PAH line, the error is just the centroid uncertainty of that line. 
The deviation in the redshift from each line is calculated as $1\sigma=\sqrt{\sum_{i}(z_{i}-\langle z\rangle )^{2}}/\sqrt{n-1}$. 
We verified the centroid uncertainties by performing Monte Carlo simulations. We take a Gaussian of the same height and width as each PAH feature and we run 1000 trials where we add noise which has the same standard deviation as the noise in the real data. For each trial, we fit the data to get the line center and keep track of the difference between the input line center and the measured line center. The centroid error is estimated from the standard deviation of this difference for all 1000 trials. We repeat this for each source. 
With the exception of GN26 whose spectrum has a very high SNR, the centroid uncertainties for individual PAH features range from 0.08--0.25$\,\mu$m which corresponds to 0.5--1.5 pixels. This is roughly an order of magnitude higher than what we would expect from the instrument based on the pixel size in the case of no noise. The total error in the IRS redshift ranges from 0.01--0.05 for these SMGs. 

\clearpage
 \begin{figure*}
\plotone{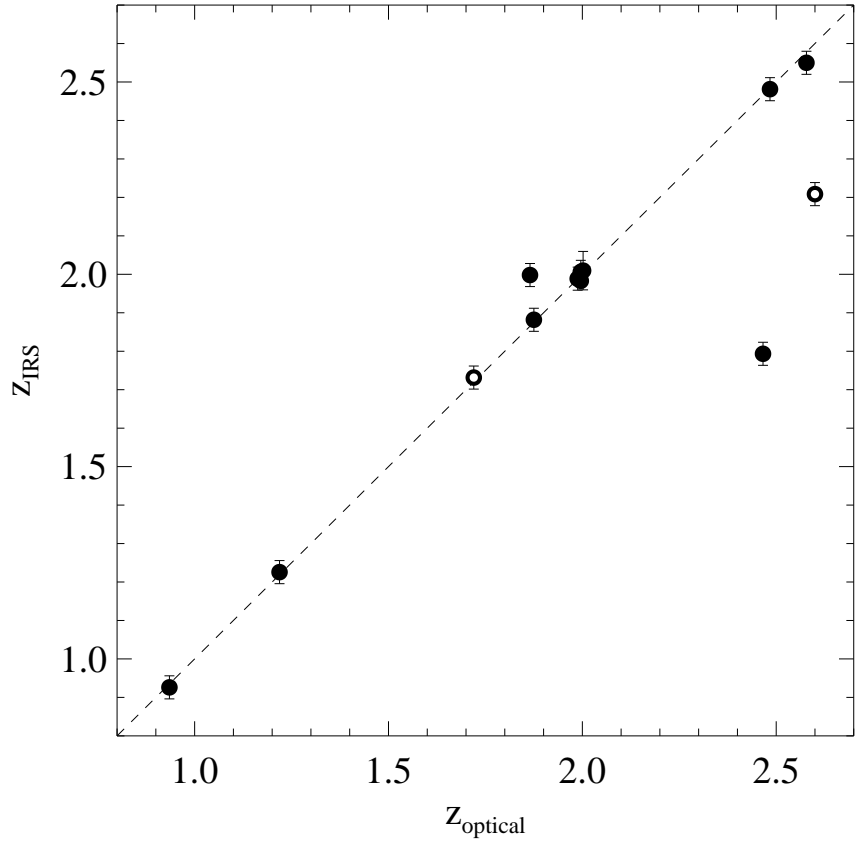}
\caption{Comparison of IRS spectroscopic redshifts and previous optical redshifts. All optical redshifts are spectroscopic expect for the two sources indicated by open symbols. There appear to be 2 optical spectroscopic redshifts which disagree with our new IRS redshifts. In these cases we believe that the IRS redshift is correct for the SMG, since the mid-IR emission is more likely to trace the submm emission,
than the optical/UV emission. 
\label{fig:redshift}}
\end{figure*}
\clearpage

Table \ref{tab:redshift} lists the new IRS redshifts and the previous optical, spectroscopic or photometric, redshifts. The new IRS mid-IR redshifts are compared with the optical redshifts in Fig.~\ref{fig:redshift}.
Two of our 13 sources seem to have very different redshifts as determined from IRS and optical spectroscopy (C2 and GN06)
which we discuss in Appendix \ref{ap:sources}).
Since the mid-IR and submm wavelengths both trace dust emission, we believe that the IRS redshifts presented here are correct for the galaxies which are causing the submm emission.

\subsection{Spectral decomposition of SMG composite spectrum}
\label{sec:comp}
\begin{figure*}
\plottwo{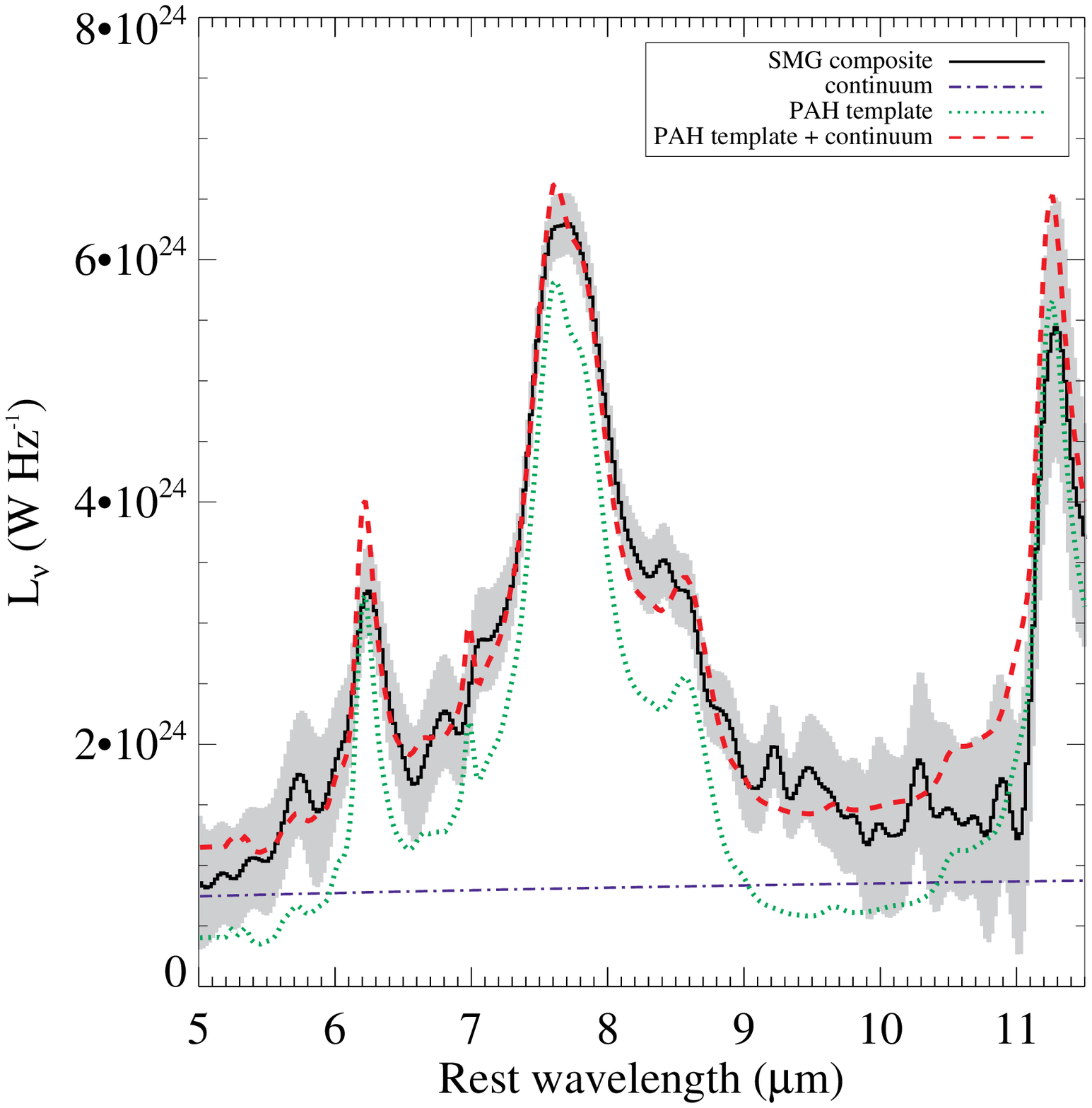}{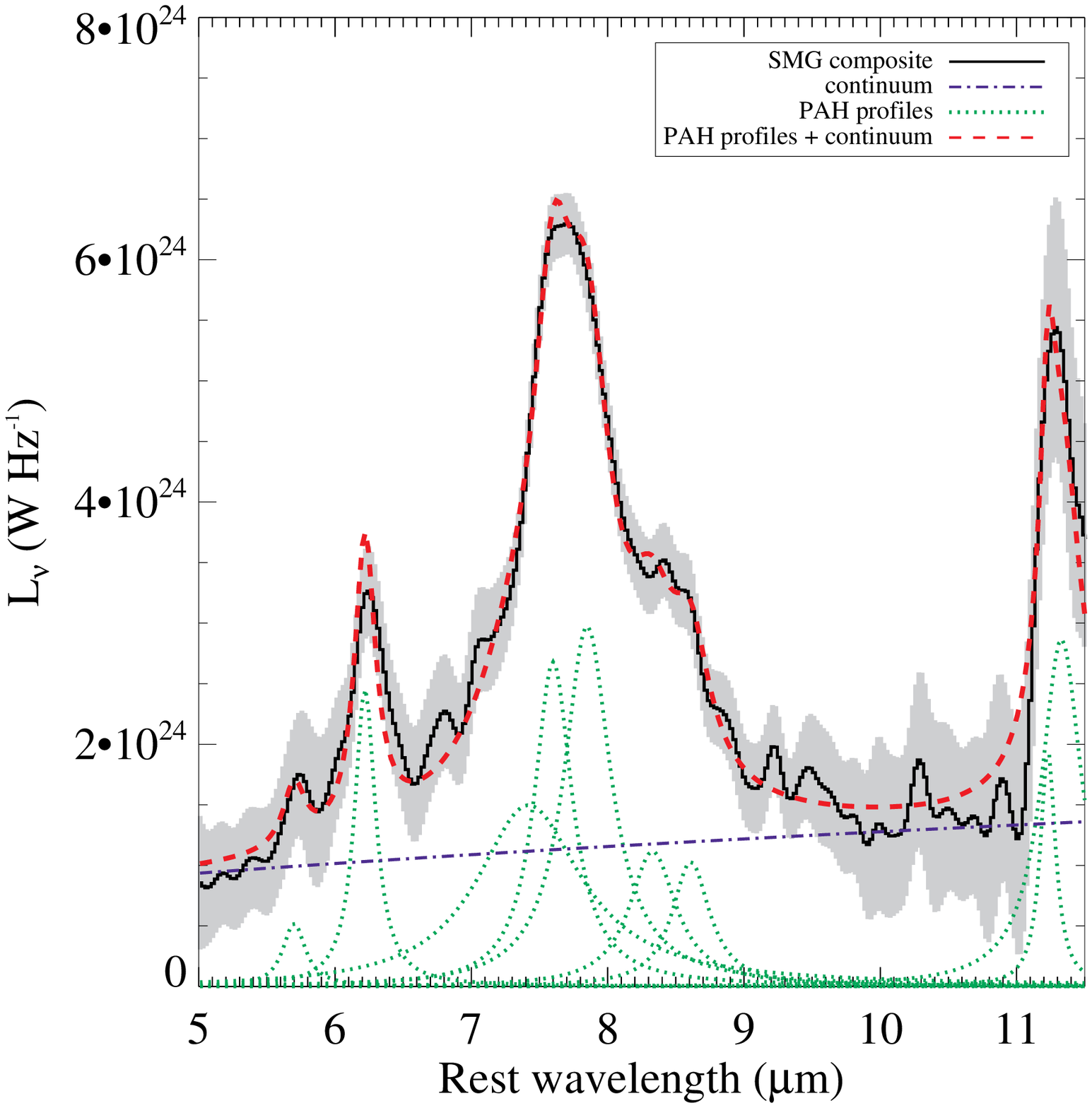}
\caption{Composite IRS spectrum of 12 SMGs (excluding C1). This SMG composite spectrum (black curve) has been smoothed by a Gaussian with a full width at half maximum (FWHM) of 0.12$\,\mu$m (rest-frame) which corresponds to the average resolution of the IRS.
The shaded region shows the 1$\sigma$ uncertainty. 
Left panel: The red curve is the best-fit model (Equation \ref{equ:model}) which is composed of a PAH template (green curve) and a power-law continuum component (blue curve), each with extinction applied to them. 
Right panel: The composite is fit using individual PAH line profiles and a continuum, each with extinction. 
In either case, we find that the PAH emission dominates the mid-IR luminosity.
\label{fig:comp}}
\end{figure*}

We can study the ensemble properties of the SMG sample by averaging together the individual mid-IR spectra.
When plotted in rest-frame luminosity units, the spectra for all SMGs in this sample show very little dynamic range. Therefore we are able to average the spectra without having to normalize each at a specific wavelength which can introduce additional biases in the composite spectrum. 
We exclude C1 from the composite spectrum, since it shows a very different spectral shape from all the other SMGs (see Section \ref{ap:sources}).
The error on the composite spectrum is calculated from combining the errors on the individual flux measurements. 
We restrict the final SMG composite spectrum to areas where there are $>3$ data files present at that wavelength which results in a wavelength coverage of $\sim$5--12$\,\mu$m in the rest frame. 
Fig.~\ref{fig:comp} shows our composite spectrum of 12 SMGs calculated from averaging the individual spectra (excluding C1).
\clearpage
\clearpage
We detect the 6.2, 7.7 and 11.3$\,\mu$m emission lines, with a hint of the 8.6$\,\mu$m emission line in the composite spectrum. The composite spectrum also shows a very flat underlying continuum.
In order to quantify the fractional contribution of star-formation and AGN activity to the emission at these wavelengths, we fit our composite spectrum with a model containing the 3 components discussed in Section \ref{sec:intro}. The PAH emission is fit using two different templates: 1) the mid-IR spectrum of M82 (F{\"o}rster Schreiber et al.~2003) and, 2) the starburst composite template from Brandl et al.~(2006). M82 is a prototypical star forming galaxy and its mid-IR spectrum is expected to be dominated by emission from star-forming regions. The starburst template from Brandl et al.~(2006) is a composite of 13 local starburst galaxies with bright flux densities and without a strong AGN component. As we show below, these templates are composed almost entirely of the sum of the individual PAH features. 
The AGN emission is characterized by a power law with both the normalization and slope as free parameters. For the extinction, we obtain $\tau_{\nu}$ from the Draine (2003) extinction curves and applied it to both the PAH and continuum components separately. The extinction curve is not just monotonic in wavelength and contains silicate absorption features, the most notable being at 9.7$\,\mu$m.
Our model, $F_{\nu}$, can be expressed as 

\begin{equation}
\label{equ:model}
F_{\nu}=c_{1}\,\nu ^{-c_{2}} e^{-c_{3}\,\tau_{\nu}} +c_{4}\,\rm{f_{\nu}} e^{-c_{5}\,\tau_{\nu}} 
\end{equation}

\noindent where $f_{\nu}$ is the PAH template. We performed a $\chi^{2}$ minimization fit for each of the two PAH templates and solved simultaneously for $c_{1}$, $c_{2}$, $c_{3}$, $c_{4}$ and $c_{5}$.

The best fit model, $F_{\nu}$, is shown in the left panel of Fig.~\ref{fig:comp} as the red dashed curve. The blue dash-dot and green dotted curves show the individual contributions from the continuum and PAH components, respectively. 
The best-fit PAH template was the starburst composite from Brandl et al.~(2006) although M82 gave very similar results. With this PAH template, the extinction needed is $\tau_{9.7}\sim1$.
The slope of the power law is very shallow ($c_{2}\sim0.2$). 
The continuum component accounts for 30\% of the luminosity at these wavelengths (5--11.5$\,\mu$m). This implies that the mid-IR luminosity of SMGs is dominated by star formation with a maximum mid-IR AGN contribution of 30\%.

To be clear about the origin of the emission in the PAH template, we also fit the SMG composite spectrum to individual PAH line profiles and a continuum with extinction. 
We include the PAH lines at 5.70, 6.22, 7.42, 7.60, 7.85, 8.33, 8.61, 11.23, 11.33$\,\mu$m, each modeled as a Drude profile (e.g.~Draine \& Li 2007) where the height of each line is treated as a free parameter. The resulting best-fit model shown in the right panel of Fig.~\ref{fig:comp} is very similar to the best-fit model which includes a PAH template and continuum (left panel). 
\clearpage
\begin{figure*}
\plotone{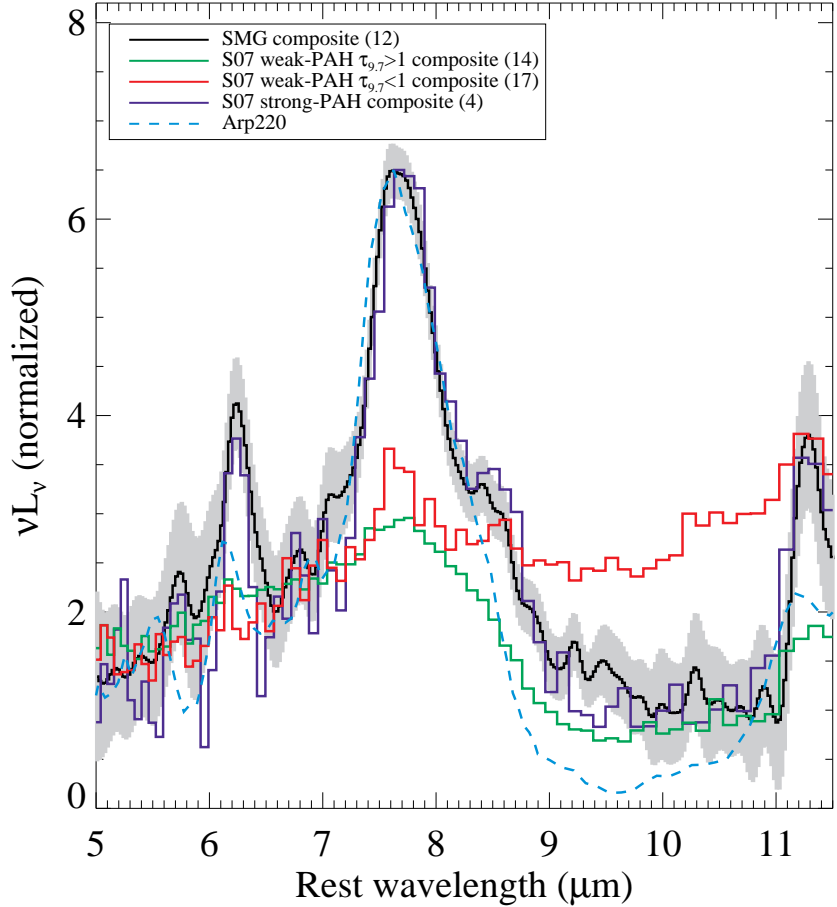}
\caption{SMG composite spectrum compared to those other ULIRGs. The SMG composite spectrum is the black curve and the shaded region shows the 1$\sigma$ error. The colored histograms are high redshift {\it Spitzer} 24\,$\mu$m-selected ULIRGs observed with the IRS from Sajina et al.~(2007, S07); the green and red curves are composite spectra of the weak-PAH sources for $\tau_{9.7}>1$ and $\tau_{9.7}<1$, respectively, and the blue curve is the composite spectrum for the strong-PAH sources. The light blue curve is the local ULIRG Arp220 (Charmandaris et al.~1999; Fischer et al.~1999). 
All curves have been normalized at $\sim7\,\mu$m.
The numbers in the legend in brackets indicate the number of sources in each composite spectrum. 
\label{fig:comp2}}
\end{figure*}
\clearpage
In Fig.~\ref{fig:comp2}, we show the SMG IRS composite spectrum compared to the IRS spectra of $z\sim2$ {\it Spitzer} 24\,$\mu$m-selected ULIRGs from Sajina et al.~(2007) and the local ULIRG Arp220. While Arp220 is often considered the typical local analogue to high redshift SMGs, it appears to have more silicate absorption and less 6.2 and 11.3$\,\mu$m PAH emission than the average SMG. Recall that the sample in Sajina et al.~(2007) consists of red 24$\,\mu$m-selected galaxies with $S_{24}>0.9\,$mJy, which is much brighter than most of the SMG population. The majority of the ULIRGs in the {\it Spitzer}-selected sample have weak PAH emission and are AGN-dominated; the green and red curves in Fig.~\ref{fig:comp2} are composite spectra of the weak-PAH (i.e.~AGN-dominated) {\it Spitzer} 24\,$\mu$m-selected ULIRGs with $\tau_{9.7}>1$ and $\tau_{9.7}<1$, respectively. 
The blue curve is the composite spectrum for the strong-PAH {\it Spitzer} 24\,$\mu$m-selected ULIRGs and is very similar to the SMG composite. In particular the relative line strengths are consistent. The strong-PAH {\it Spitzer}-selected sources represent only a small fraction of the Sajina et al.~(2007) sample and these sources tend to have a higher detection rate in the (sub)mm than the rest of their sample (Sajina et al.~in preparation). 
From Yan et al.~(2007) and Sajina et al.~(2007), the surface density of the $z\sim2$ {\it Spitzer} 24\,$\mu$m-selected ULIRGs is 10.4 deg$^{-2}$ and the strong-PAH sub-sample of these sources has a surface density of 2.6 deg$^{-2}$.
The surface density of SMGs with $S_{850}>4\,$mJy (i.e.~$L_{\rm{IR}}\gtrsim 5\times10^{12}\rm{L_{\odot}}$, Pope et al.~2006) is 844 deg$^{-2}$ (Coppin et al.~2006).
While most submm-selected galaxies are much fainter at $24\,\mu$m, the strong-PAH {\it Spitzer} 24$\mu$m-selected ULIRGs appear to be a very small sub-sample of the SMG population.

\subsection{Spectral decomposition of individual galaxies}
\label{sec:specfit}
Given the good agreement between our SMG composite spectrum and the spectrum of starburst galaxies, we 
proceed to fit the individual SMG mid-IR spectra to the model given in Equation \ref{equ:model}. 
For each of the 13 SMGs we solve for the five parameters in Equation \ref{equ:model} and plot the best-fit model in Fig.~\ref{fig:spec} as the red dashed curve. The separate starburst and AGN components including extinction are shown as the green and blue curves. For most SMGs, the continuum component is zero and not plotted in Fig.~\ref{fig:spec}.
We note that the spectrum of GN26 is even lower than the model at long wavelengths which confirms that the AGN component must be negligible in this source. 

From our model fits, we derived the fraction of the mid-IR luminosity which comes from the continuum component. 
The continuum component provides an upper limit to the AGN contribution since hot dust present in the most energetic HII regions can also contribute to the continuum. When the best-fit model does not include a continuum component, we calculate an upper limit by conservatively assuming that all of the emission around $6\,\mu$m is coming from a continuum component.  
We used this fraction to classify the mid-IR spectrum and the results are listed in Table \ref{tab:agn}. $8/13$ SMGs are clearly starburst-dominated sources (continuum component $\lesssim50\%$ of the mid-IR luminosity) and only $2/13$ SMGs are AGN-dominated (continuum component $\gtrsim50\%$ of the mid-IR luminosity) in the mid-IR. The remaining sources have a combination of PAH and continuum emission in the mid-IR.
The results for the individual galaxies are in agreement with the composite spectrum which showed a 30\% AGN contribution on average.

\subsection{AGN classification}
Alexander et al.~(2005) showed that deep X-ray observations are a powerful tool for identifying the presence of AGN in SMGs. 
They conclude that the majority of SMGs host an AGN, but that the bolometric output from SMGs is dominated by star formation because of the low ratio of X-ray luminosity to IR luminosity. They also stress that some fraction of SMGs will be AGN-dominated. 
Following the same method as Alexander et al.~(2005) we list the X-ray classification in Table \ref{tab:agn} where a hard X-ray detection tells us that the X-ray emission is dominated by AGN activity. 

We find that the X-ray classification gives a higher fraction of AGN sources ($6/13$) than the mid-IR spectra ($2/13$) for SMGs with $S_{24}>200\,\mu$Jy.
For the X-ray classified AGN sources, the X-ray emission expected from star formation (e.g. Bauer et al. 2002) is less than that observed indicating that there is an AGN present. Our IRS observations show that this AGN is not important to the mid-IR luminosity.
On the other hand, the line widths and line ratios in the UV/optical spectra more often indicate that these SMGs are starburst sources (Swinbank et al.~2004; Chapman et al.~2005) and appear to be insensitive to the AGN activity in many sources.

C1 is classified as a starburst from the X-ray data but shows a very strong AGN in the mid-IR. This SMG is likely to host a very obscured AGN, the kind of system that can help explain the unresolved portion of the hard X-ray background (e.g.~Brandt \& Hasinger 2005; Daddi et al.~2007). 
See Appendix \ref{ap:sources} for more discussion about C1. Combining the X-ray imaging with mid-IR spectroscopy provides the best census 
of AGN in SMGs. 

In Pope et al.~(2006), we found that most SMGs showed evidence for a stellar bump in the 4 IRAC channels. 
We found only 4 sources which had a good fit to a simple power-law from IRAC+24$\,\mu$m photometry. Only one of these is in our IRS sample: GN04, since the others are too faint for spectroscopy
at 24$\,\mu$m. We find that the IRAC SED, the mid-IR spectrum and X-ray classification all
converge on the conclusion that there is an energetically important AGN in GN04. In contrast,
the UV/optical spectra of this SMG did not reveal the presence of an AGN in it. 
\clearpage
\begin{figure*}
\epsscale{.80}
\plotone{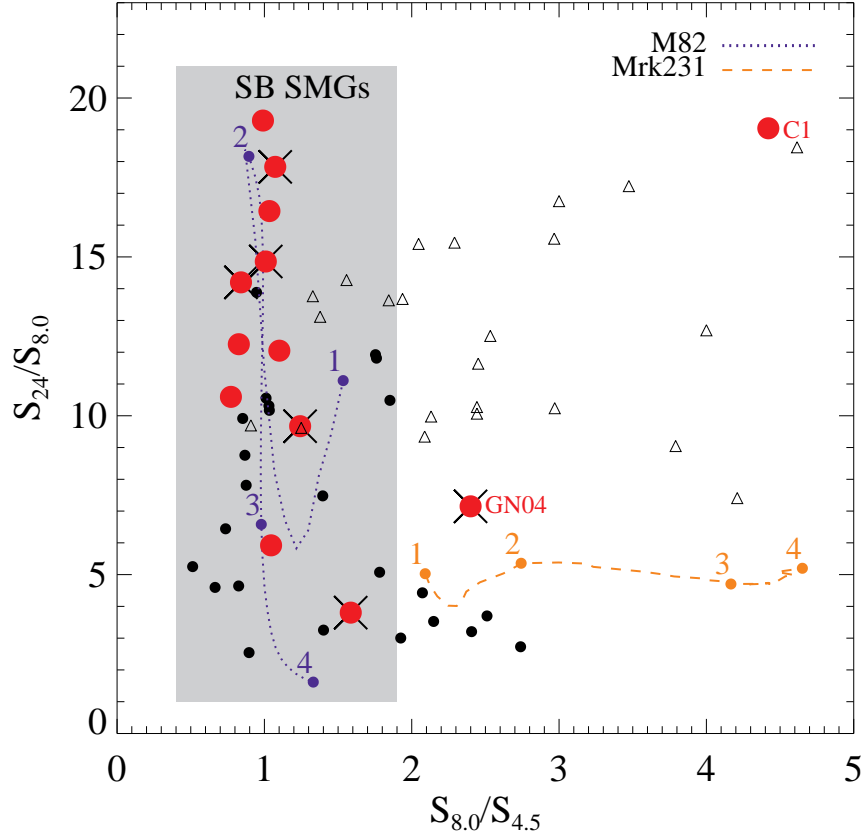}
\caption{{\it Spitzer} color-color diagram as an AGN diagnostic (See also Ivison et al.~2004). The orange dashed and blue dotted curves show the positions of Mrk231 (an AGN) and M82 (a starburst), respectively, as a function of redshift (redshift is indicated by the numbers along the track). 
The large red dots are the SMGs from this study for which we have IRS spectra. 
The 2 SMGs (C1 and GN04) which have a $>50\%$ contribution from continuum (AGN) emission to the mid-IR luminosity based on the IRS spectra (see Table \ref{tab:agn}) both lie off the M82 starburst sequence.  
Based on our IRS results, SMG which lie in the light shaded region are starburst (SB)-dominated in the mid-IR. 
The crosses indicate sources which are classified as X-ray AGN (Alexander et al.~2005). The smaller dots are the rest of the SMGs from Pope et al.~(2006). Using our IRS results, the fraction of all SMGs which are starburst-dominated in the mid-IR (i.e.~are found in the shaded region)  is $\gtrsim80\%$. 
The small open triangles are the {\it Spitzer} 24\,$\mu$m-selected ULIRGs from Sajina et al.~(2007) with $z>1$; most of these sources are outside the starburst region defined for SMGs. 
\label{fig:cc}}
\end{figure*}
\clearpage

Another approach for separating sources which contain a mid-IR AGN component is through the use of {\it Spitzer} color-color diagrams. 
Ivison et al.~(2004) proposed that the $S_{8}/S_{4.5}$ vs $S_{24}/S_{8}$ color plane could be used to identify AGN-dominated SMGs. 
The IRS spectra have allowed us to decompose the starburst and AGN components in SMGs therefore we are in a position to test the plot as a diagnostic.
In Fig.~\ref{fig:cc}, we plot this {\it Spitzer} color-color diagram. The orange dashed and blue dotted curves show the positions of Mrk231 (mid-IR spectrum from Rigopoulou et al.~(1999) spliced with a fit to the {\it Infrared Astronomical Satellite}, {\it IRAS}, photometry) and M82 (F{\"o}rster Schreiber et al.~2003), respectively, as a function of redshift (redshift is indicated by the numbers along the track). Based on these broad-band colors, AGN-dominated sources are expected to lie along the Mrk231 tracks while starburst-dominated systems should have similar colors to M82. 
Sajina, Lacy \& Scott (2005) showed that while AGN-dominated sources do lie on this AGN sequence, there are also just as many high redshift PAH-dominated sources with high extinction in this region. 
The large red dots are the SMGs from this study for which we have IRS spectra.
Interestingly, both SMGs (C1 and GN04) which have a $>50\%$ contribution from continuum (AGN) emission to the mid-IR luminosity based on the IRS spectra (see Table \ref{tab:agn}) lie off the starburst sequence. 
Based on our IRS results, SMG which lie in the light shaded region of Fig.~\ref{fig:cc} are starburst-dominated in the mid-IR.
SMGs outside the shaded region likely have a combination of starburst and AGN emission in the mid-IR or lie at the highest redshifts ($z\gtrsim4$).

This division between starburst and AGN-dominated sources based on IRS spectra agrees with the broad-band colors. The crosses indicate sources which are classified as X-ray AGN (Alexander et al.~2005). Many of these lie along the starburst sequence and have IRS spectra which are starburst-dominated. While the X-ray emission may be telling us that an AGN is present in the system, it does not seem to be a very good diagnostic of whether the AGN contributes to the bolometric luminosity. 
Sources which have AGN-dominated mid-IR spectra, like C1, clearly show very extreme colors in this diagram however there are not enough of these sources to characterize their allowable positions in this color-color space. 
The smaller dots are the rest of the SMGs from the GOODS-N submm super-map from Pope et al.~(2006). 
Extrapolating our IRS results, the fraction of all SMGs which are starburst-dominated in the mid-IR is $\sim80\%$. 
The small open triangles are the $z>1$ bright 24$\mu$m ULIRGs from Sajina et al.~(2007); very few of these sources lie within the starburst region defined for SMGs indicating that they have a wider range of mid-IR spectra.

\section{Full SED fits}
\label{sec:sed}
In Pope et al.~(2006), we fit the 24$\,\mu$m, 850$\,\mu$m and radio photometry to a suite of models to determine the total infrared luminosity and dust temperatures. We had
used the Chary \& Elbaz (2001, hereafter CE01) templates and modified them by adding additional extinction from the Draine (2003) models. 
The CE01 templates were developed to be representative of local galaxies therefore they contain an inherent luminosity-temperature (L-T) relation.
When fitting, we had allowed the models to scale so that we could solve for both the luminosity and the temperature independently (i.e.~a model with high luminosity but cool average dust temperature is allowed). We had found that SMGs fit best to scaled-up versions of lower luminosity templates which had cooler average dust temperatures than local ULIRGs. In those fits the short wavelength part of the spectrum was constrained only by the 24$\,\mu$m flux. 
From the present study it is apparent that the 24$\,\mu$m flux of SMGs will vary considerably with redshift, due to the presence of the
PAH lines. Therefore, we have repeated the full SED fitting now that we have the mid-IR spectra.
\clearpage
\begin{figure*}
\plotone{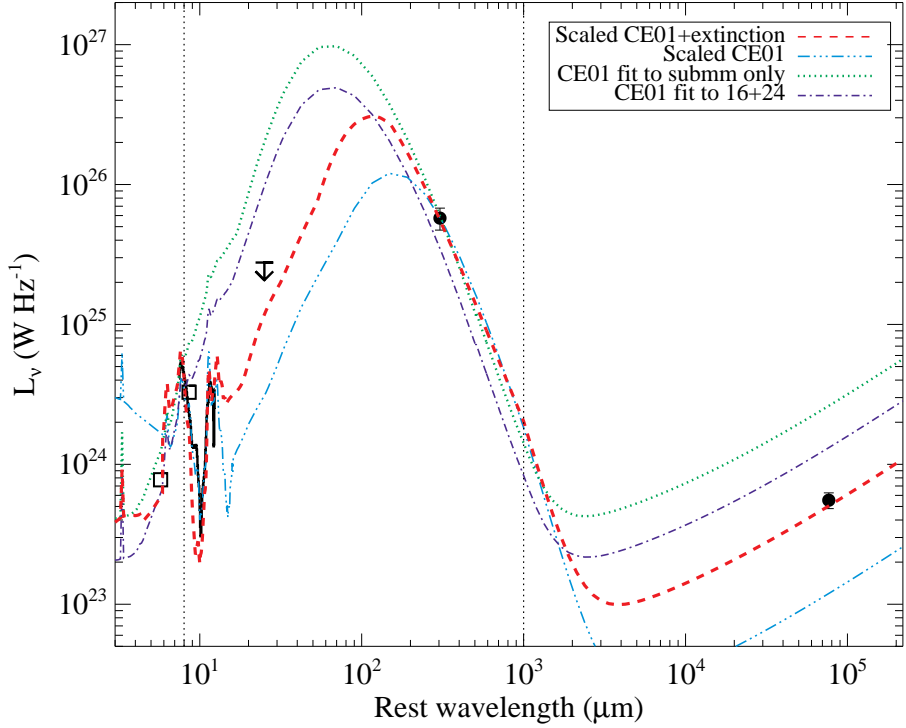}
\vspace{0.5in}
\caption{IR SED of source C2. This figure demonstrates our template fitting to the full SED from mid-IR to radio wavelengths. 
The green dotted curve is the CE01 template which best matches the submm flux while the dot-dash dark blue curve is the CE01 template which best matches the 16 and 24$\,\mu$m fluxes. 
Both of these models are clearly poor fits to the other data points and have a higher total infrared luminosity than the best-fit to all the available data. The light blue dash-dot-dot curve is the best fit to all the data points $>7\,\mu$m rest frame where we have allowed the CE01 templates to scale. While this provides a good fit to the mid-IR spectrum and the submm flux, the radio flux and the total infrared luminosity are underestimated.
The red dashed curve is a scaled CE01 template with additional extinction from the Draine (2003) extinction curve which provides the best-fit to our data. We require both a scaling of local templates (in this case CE01) and additional extinction to fit the SED of high redshift SMGs. 
\label{fig:C2}}
\end{figure*}
\clearpage

Fig.~\ref{fig:C2} demonstrates why we need to allow the CE01 templates to scale (i.e.~luminosity and temperature are both variables) and the need for an additional extinction term. 
Here we plot the mid-IR to radio SED for one of our SMG sources, C2. 
First we simply plot the CE01 template which best matches the submm flux only as the green dotted curve and the CE01 template which best matches the 16 and 24$\,\mu$m fluxes as the dot-dash dark blue curve. 
For these two models we emphasize that the CE01 templates have not been allowed to scale therefore they follow the local luminosity-temperature relation. Because SMGs are so luminous at mid-IR and submm wavelengths, the green and dark blue models have a warm average dust temperature ($\sim50$K).
The light blue dash-dot-dot curve is the best fit to all the photometry between rest-frame
$7-300\,\mu$m where we have allowed the CE01 templates to scale. While this provides a good fit to the mid-IR spectrum and the submm flux, the radio flux and $L_{\rm{IR}}$ are underestimated
compared to our best fit. We know that the CE01 templates need to be scaled up to match the photometry of SMGs however as we scale them up we are introducing a greater dust mass which must then cause more extinction.
The red dashed curve is a scaled CE01 template with additional extinction from the Draine (2003) extinction curve which provides the best-fit to our data. We require both a scaling of local templates (in this case CE01) and additional extinction to fit the SED of high redshift SMGs. 

\begin{figure*}
\epsscale{.80}
\plotone{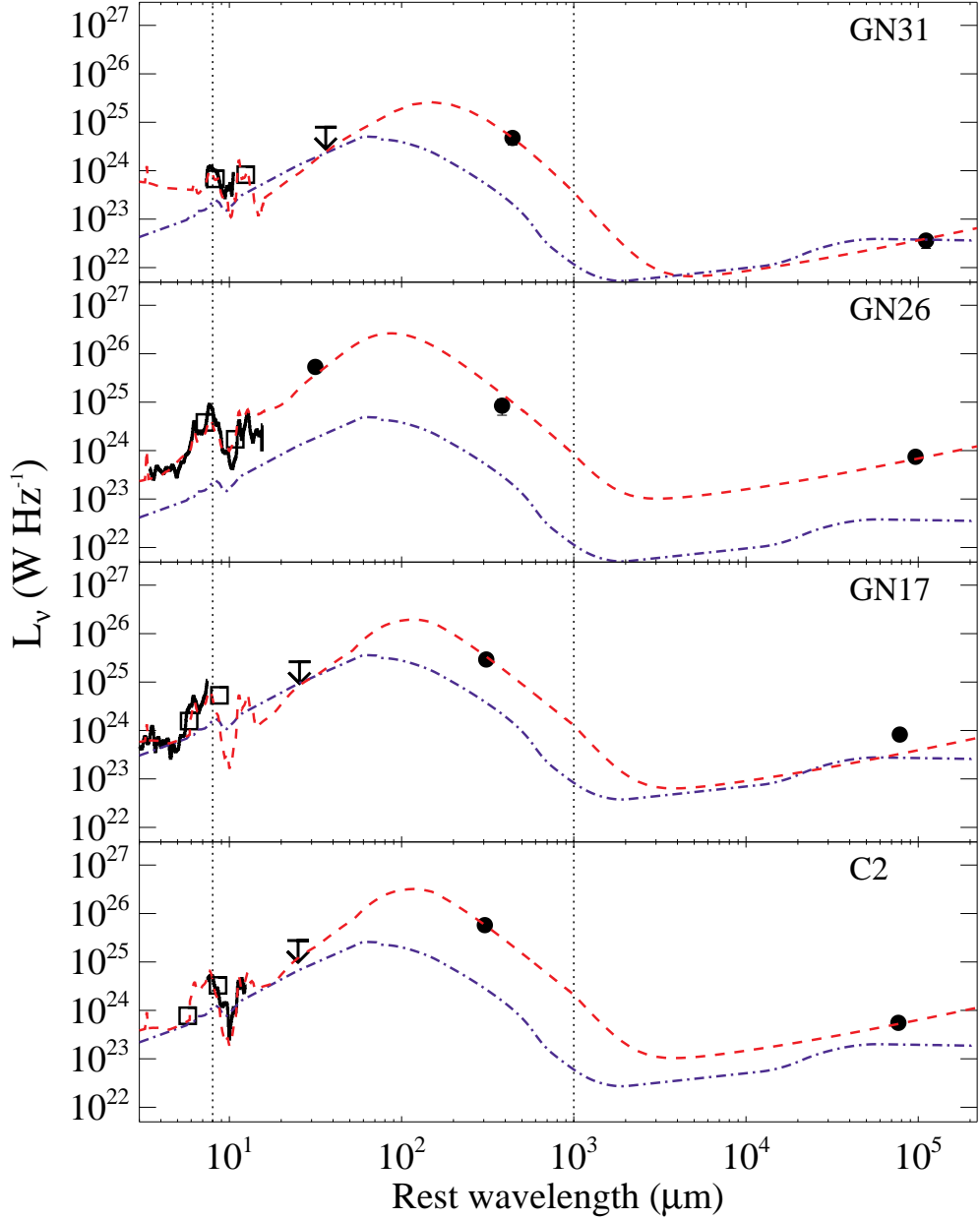}
\vspace{0.5in}
\caption{Mid-IR to radio SEDs of SMGs. The IRS spectrum (dark curve) plus 70$\,\mu$m, submm and radio photometry (solid circles) are used in the fit. The open squares indicate the photometry at 16 and 24$\,\mu$m. The red dashed line is the best fit modified CE01 template while the blue dash-dot curve is Mrk231 normalized to the AGN fraction of the mid-IR given in Table \ref{tab:agn}.  We used the latter to determine the contribution of the AGN component to $L_{\rm{IR}}$.
The horizontal dotted lines indicate the region within which $L_{\rm{IR}}$ is calculated.
\label{fig:sed}}
\end{figure*}

We thus fit the IRS spectrum, the 70$\,\mu$m, submm and radio photometry to the CE01 templates plus Draine (2003) extinction for each of the 13 SMGs in this paper\footnote{For the GN39 double system, we split the total submm flux according to the radio flux of each component (note that this is the same as splitting by 24$\,\mu$m flux).}. 
We solve for the three free parameters (luminosity, average dust temperature and extinction) and find the template with the lowest $\chi^{2}$. 
Fig.~\ref{fig:sed} shows the full SED fits. In general, the resulting fits are very good. The dark curve and dots are the points used in the fit, while the open symbols show the other photometry points (mostly 16 and 24$\,\mu$m) which were not used in the fits but agree well with the best-fit SED. We required a scaling of 10--100 in luminosity to fit the local CE01 templates to the SMG photometry. Since high redshift SMGs can be well fit by modified CE01 templates which contain an inherent radio-IR correlation, this suggests that SMGs also follow a similar radio-IR correlation to local galaxies (see also Kovacs et al.~2006). 

We integrated the full SEDs in Fig.~\ref{fig:sed} from 8--1000$\,\mu$m to get the total infrared luminosity. Values of $L_{\rm{IR}}$ are listed in Table \ref{tab:lines} and they are in good agreement with the values from Pope et al.~(2006). 

\clearpage
\clearpage
{\plotone{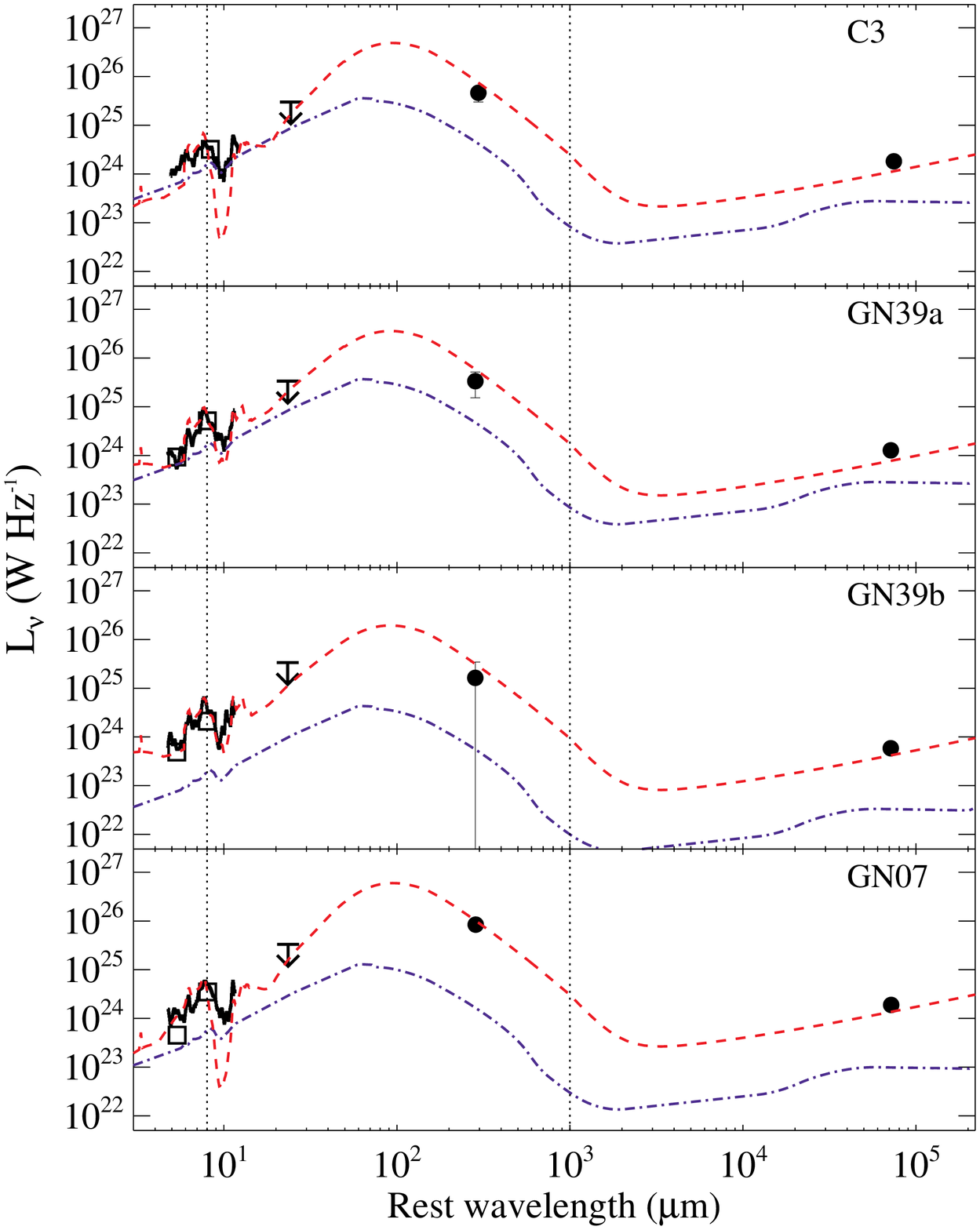}}\\[5mm]
\centerline{Fig. 9. --- Continued.}
\clearpage
\thispagestyle{empty}
\vspace*{-20mm}
{\plotone{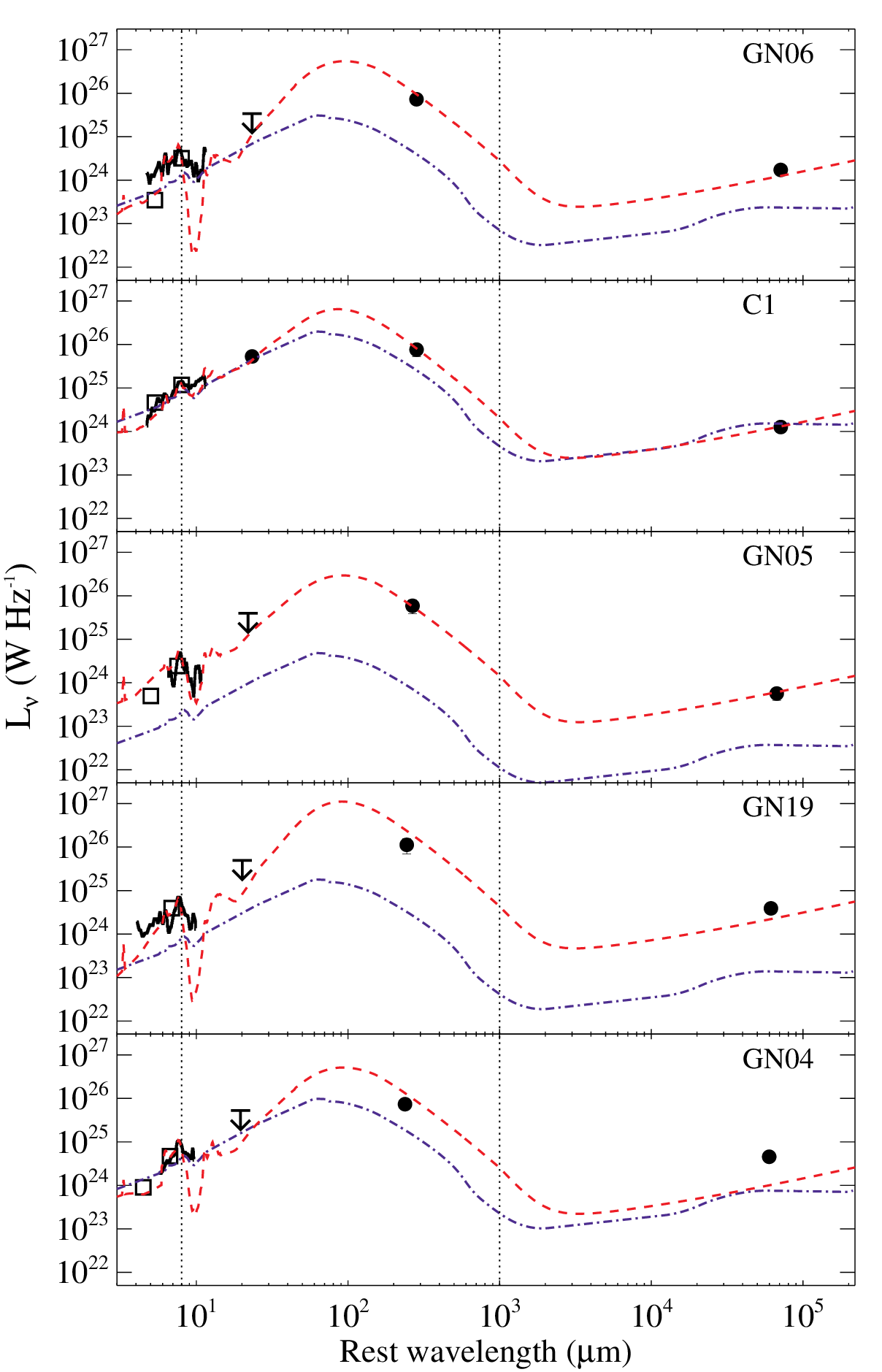}}\\[5mm]
\centerline{Fig. 9. --- Continued.}
\clearpage

By simply scaling to the wavelength where the SED is a maximum as an estimate of the average dust temperature, we obtain a median of 32$\,$K for these 13 SMGs. This is consistent with previous estimates of dust temperatures in SMGs (Chapman et al.~2005; Kovacs et al.~2006; Pope et al.~2006). We also found a similar trend between luminosity and temperature as found in Chapman et al.~(2005) which is offset to cooler temperatures from the local correlation. 
Knowing the shape of the far-IR SED and the total luminosity, we can also estimate the dust mass -- following the relation given in Dunne et al.~(2000), the median dust mass for this sample is $2.4\times10^{8}$M$_{\odot}$. This is roughly an order of magnitude higher than dust masses observed in galaxies locally although the local galaxies studied are generally much less luminous.

In order to quantify the effect of the extinction term on the models, we calculate $L_{\rm{IR}}$ without the extinction. The median amount that the extinction term changes $L_{\rm{IR}}$ by is 13\% (0--20\%), and so models which do not include the extinction term will incorrectly determine $L_{\rm{IR}}$ by around 13\%. We find an average value of $\tau_{9.7}\sim1$ and we don't see any correlations between the extinction and $L_{\rm{IR}}$ for the SMGs given the error bars. 

For sources which we found to contain an AGN component in the mid-IR, we calculate the contribution to $L_{IR}$ from the AGN.
Using {\it ISO} spectra, Tran et al.~(2001) found that the ratio of $L_{5-10\,\mu m}$/$L_{\rm{IR}}$ was about 2.5 times greater for AGN than for starbursts i.e. as you extrapolate from mid-IR to far-IR wavelengths AGN become less important bolometrically than starbursts. 
In order to quantify this for our sample, we scale the Mrk231 (an AGN whose $L_{\rm{IR}}$ is known to be dominated by AGN emission and not star formation, Armus et al.~2007) template to the AGN fraction (listed in Table \ref{tab:agn}) of the mid-IR luminosity and then integrate the SED from 8--1000$\,\mu$m to get the contribution to $L_{IR}$ from the AGN component. For sources which have an upper limit to the AGN contribution, we plot the scaled Mrk231 in Fig.~\ref{fig:sed} at this limit. The values of $L_{IR}^{AGN}$ from integrating under the curves and $L_{IR}^{SB}=L_{IR}-L_{IR}^{AGN}$  are given in Table \ref{tab:lines}.
As seen in Fig.~\ref{fig:sed}, only C1 has enough AGN contribution in the mid-IR to make a significant contribution to $L_{\rm{IR}}$. Therefore the intense IR luminosities seen in SMGs are powered almost entirely by star formation.
In Section \ref{sec:dis}, we discuss how this luminosity can be converted to give us the star formation rate (SFR).

\section{PAH luminosities}
\label{sec:lpah}

PAH line profiles have broad wings which contribute a significant fraction of the line flux. 
PAH line fluxes therefore, need to be consistently measured in a uniform way before comparisons between local sources and high redshift sources can be performed. 
We have measured line fluxes of all sources consistently, including the comparison sample of low redshift starbursts (Brandl et al.~2006), using the technique described in Section \ref{ap:meas}.

The PAH line luminosities and equivalent widths of the SMGs including $3\sigma$ upper limits when the line is undetected are listed in Table \ref{tab:lines}. The 6.2$\,\mu$m line is the cleanest to measure and use for diagnostics since it is the least affected by silicate extinction and line blending. We can see that $L_{\rm{PAH,}7.7}$ for the SMGs varies by over an order of  magnitude, while $L_{\rm{PAH,}6.2}$ and $L_{\rm{PAH,}11.3}$ have much smaller dynamic ranges of a factor of a few. While this might seem odd, we note that only a subset (40--60\%) of our sample have 6.2 and 11.3$\,\mu$m spectral coverage whereas 92\% have coverage of the 7.7$\,\mu$m emission line. Furthermore, the redshift range for the sources with 6.2 and 11.3$\,\mu$m spectral coverage is very small and therefore we expect a small dynamic range in PAH luminosity.
The sources which are dominated by an AGN in the mid-IR, as classified by our IRS spectra (C1 and GN04), both show suppressed 6.2$\,\mu$m emission relative to their 7.7$\,\mu$m emission. 

\begin{figure*}
\plotone{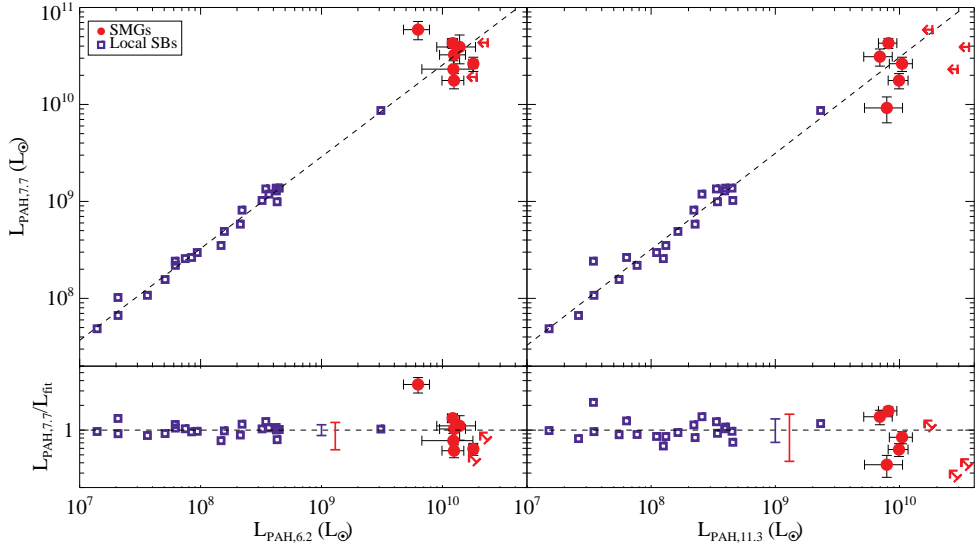}
\vspace{0.1in}
\caption{Plot comparing the PAH line luminosities
of SMGs presented in this paper (red circles) to that of local starbursts (blue squares, Brandl et al.~2006). All line measurements have 
been made in a consistent way (see Section \ref{ap:meas}). $1\sigma$ error bars are shows on all points (the 10\% errors on the starbursts are similar in size to the symbols). 
The top panels show $L_{\rm{PAH,}7.7}$ as a function of $L_{\rm{PAH,}6.2}$ (left) and $L_{\rm{PAH,}11.3}$ (right) where the dashed lines shows the best fit to the starburst galaxies only.
The similar relationship seen in starbursts and SMGs between PAH line strengths, indicates that SMGs are starburst-dominated and that PAH line strengths do not vary significantly with redshift. 
The smaller bottom panels show the residual scatter after removing the fit shown in the upper panels; the larger colored error bars quantify the $1\sigma$ scatter seen in the starbursts and SMGs, respectively. 
\label{fig:lpah}}
\end{figure*}

Fig.~\ref{fig:lpah} shows the relative strengths of PAH lines compared to local starbursts (Brandl et al.~2006). 
The top panels show $L_{\rm{PAH,}7.7}$ as a function of $L_{\rm{PAH,}6.2}$ (left) and $L_{\rm{PAH,}11.3}$ (right). 
The smaller bottom panels show the residual scatter after removing the fit shown in the upper panels; the two larger colored error bars quantify the $1\sigma$ scatter seen in the starbursts and SMGs.
The outlying source in the left panel of Fig.~\ref{fig:lpah} is C1, the SMG with the largest AGN contribution. 
\clearpage
\clearpage

While the SMGs do not show enough dynamic range in PAH luminosity to pick out any strong correlations within the population, they do fall on the relation between various PAH lines established for local starburst galaxies. The dashed lines in Fig.~\ref{fig:lpah} are the best-fits to the local starbursts only (blue squares). We find essentially the same relation if we fit the local starbursts and the SMGs, indicating that the PAH luminosities in SMGs are simply scaled up (albeit by several orders of magnitude) from less luminous starburst galaxies. The best-fit relations from fitting both the SMGs and the local starburst galaxies are 

\begin{eqnarray}
\label{equ:lpah}
\rm{log}({\it L}_{\rm{PAH,}7.7})=(1.0\pm0.1)+(0.94\pm0.01) \, \rm{log}({\it L}_{\rm{PAH,}6.2})\\ 
\rm{log}({\it L}_{\rm{PAH,}7.7})=(0.5\pm0.1)+(1.00\pm0.01) \, \rm{log}({\it L}_{\rm{PAH,}11.3}).
\end{eqnarray}

The SMG C1 appears to lie off this relation with an excess of 7.7$\,\mu$m PAH emission for a given $L_{\rm{PAH,}6.2}$ or $L_{\rm{PAH,}11.3}$.
Rigopoulou et al.~(1999) found suppressed 6.2$\,\mu$m emission relative to 7.7\,$\mu$m emission in ISO spectra of local ULIRGs compared to starbursts and they conclude that this is primarily because of extinction. 
This suggests that, on average, the SMG C1 considered here may contain greater extinction or a more dominant AGN component than most SMGs and starburst galaxies since an AGN continuum will dilute the strength of the weaker 6.2 and 11.3$\,\mu$m features relative to the strong 7.7$\,\mu$m line. 
\clearpage
\begin{figure*}
\plotone{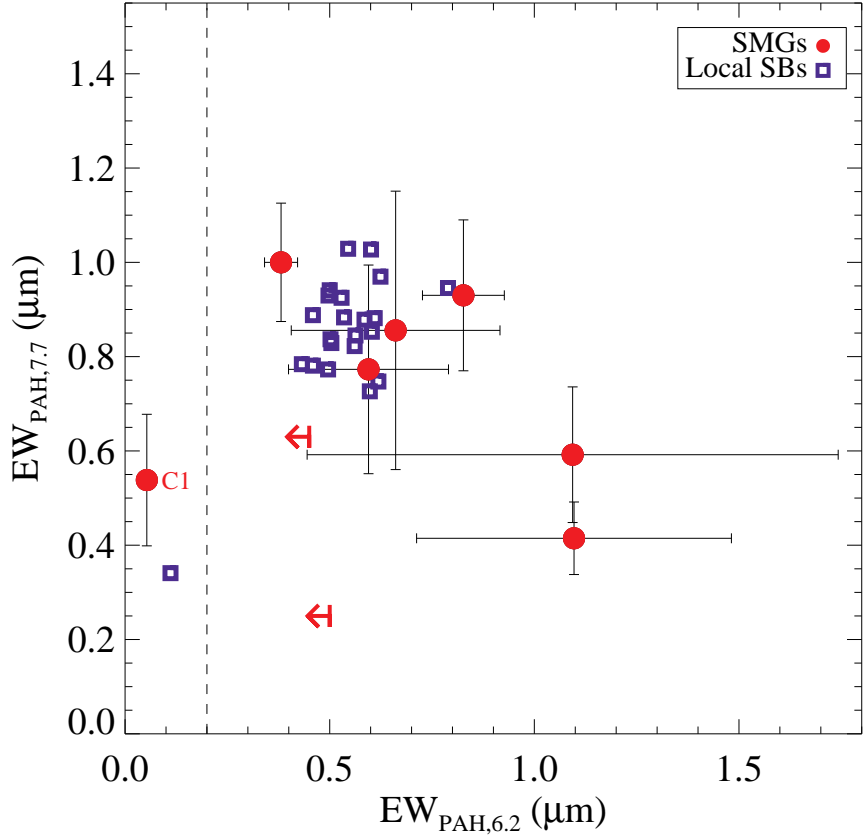}
\caption{Rest-frame equivalent widths of main PAH features. Again the SMGs are denoted by the red circles and the local starbursts (Brandl et al.~2006) by the blue squares.
All measurements have been made in a consistent way (see Section \ref{ap:meas}).
Measurements of both the 6.2 and 7.7$\,\mu$m PAH lines are not available for all SMGs. 
C1 stands out again as having a very small 6.2$\,\mu$m EW, below the values of EW$_{\rm{PAH,}6.2}=0.2$ (vertical dashed line) which is often used to separate AGN-dominated and starburst-dominated systems.
\label{fig:ew}}
\end{figure*}
\clearpage

Fig.~\ref{fig:ew} shows the equivalent widths of the 6.2 and 7.7$\,\mu$m PAH lines. The vertical dashed line at $EW_{\rm{PAH,}6.2}=0.2$ is often used to separate starburst-dominated and AGN-dominated sources (Armus et al.~2007; Sajina et al.~2007). With the exception of C1, all of our sources with measurable 6.2$\,\mu$m PAH emission have $EW_{\rm{PAH,}6.2}>0.2$, which indicates that they are starburst-dominated. 
We have again plotted the local starbursts for comparison. 
We found that the values in Brandl et al.~(2006) for EW$_{\rm{PAH,}7.7}$ were on average 1.6 times lower than our measurements of the starburst EW$_{\rm{PAH,}7.7}$, due to differences in the way the lines were measured. Therefore we have revised all the Brandl et al.~(2006) points upwards in this plot. The starbursts cover a small area of this plot, while the SMGs are spread over a larger region, although with much larger error bars. Our current data do not have sufficient SNR to determine if there is any genuine spread or correlation among the EWs, however our error bars are sufficiently small to exclude equivalent widths of less than 0.2 in either PAH line (with the exception of C1). Our SMGs show similar equivalent widths to the PAH-dominated ULIRGs from the {\it Spitzer} 24$\mu$m selected sample in Sajina et al.~(2007), if we decrease their values of $EW_{\rm{PAH,}7.7}$ by a factor of 4 to account for the different methods for measuring the EWs. With $EW_{\rm{PAH,}6.2}$ values of $\gtrsim0.5$ and $\tau_{9.7}\sim1$, high redshift SMGs lie in region 2C of the diagnostic plot in Spoon et al.~(2007). Sources in this region are classified as PAH-dominated and, not surprisingly, M82 also falls in this region. 

Utilizing {\it ISO} spectra of starbursts, ULIRGs and AGN, Genzel et al.~(1998) used the ratio of the 7.7$\,\mu$m PAH line flux to the continuum flux, $l/c$, as a diagnostic for AGN contribution in the mid-IR emission. In general, galaxies with $l/c\ge1$ are classified as starbursts and those with $l/c<1$ are AGN-dominated. 
As we stress in Section \ref{ap:meas}, this type of analysis is strongly dependent on the specific choice made for the continuum fitting procedure.
Nevertheless, we calculate $l/c\sim2.3$ for the composite SMG and we note that all SMGs in our sample satisfy the $l/c\ge1$ criterion for starburst-dominated sources. This is true even for C1 which clearly shows a strongly rising continuum indicative of the presence of an AGN.
Based on Fig.~\ref{fig:ew}, the 6.2$\,\mu$m PAH line may be more useful in weeding out AGN-dominated sources.

\begin{figure*}
\epsscale{.35}
\vspace{0.1in}
\plotone{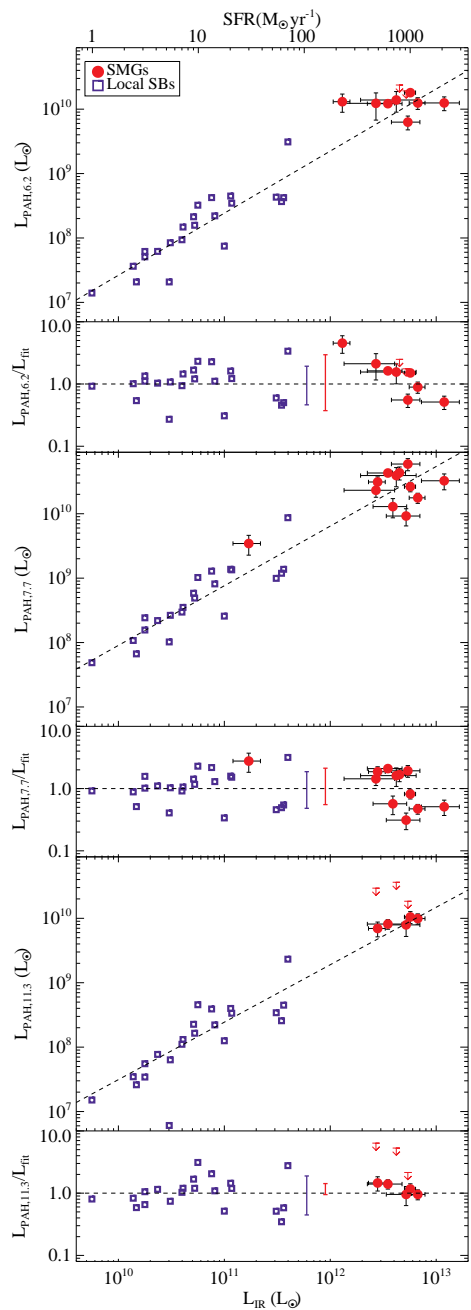}
\vspace{0.1in}
\caption{Correlations between $L_{\rm{IR}}$ and PAH luminosities. 
The dashed lines show the best-fit relations for the local starburst galaxies only; the smaller panels show the residual scatter around the best-fit relations. The SMGs clearly lie on the relations established by local starburst galaxies.
\label{fig:lir}}
\end{figure*}

In Fig.~\ref{fig:lir} we plot the PAH line luminosities as a function of $L_{\rm{IR}}$ for SMGs and local starbursts (large panels). For the SMGs, we plot the $L_{\rm{IR}}^{SB}$ coming from star formation (i.e.~after subtracting the component from the AGN). Errors in $L_{\rm{IR}}$ for the SMGs have been estimated from the error in the submm flux only, since this dominates (assuming fixed CE01 templates). 
In all panels, the dashed line is the best-fit to the local starburst galaxies; the smaller panels show the residual scatter around this best-fit relation. 
The best-fit relation for the starburst galaxies only is consistent with the fit to both the starburst galaxies and the SMGs.
The best-fit relation for both the SMGs and starburst galaxies is given by

\begin{eqnarray}
\label{equ:lir}
\rm{log}({\it L}_{\rm{PAH,}6.2})=(-2.7\pm0.1)+(1.01\pm0.01) \, \rm{log}({\it L}_{\rm{IR}})\\
\rm{log}({\it L}_{\rm{PAH,}7.7})=(-1.5\pm0.1)+(0.95\pm0.01) \, \rm{log}({\it L}_{\rm{IR}}) \\
\rm{log}({\it L}_{\rm{PAH,}11.3})=(-1.6\pm0.2)+(0.91\pm0.01) \, \rm{log}({\it L}_{\rm{IR}}).
\end{eqnarray}

The derived parameters are all within $1\sigma$ of those found by fitting only the starbursts. 
The slopes are close to one indicating a direct proportionality between $L_{IR}$ and the PAH line luminosities.

The first thing to note is that the SMGs lie on the relation established for local starburst galaxies extrapolated to very high luminosities. This is what we expect if both are dominated by the same emission mechanism, namely star formation. 
Brandl et al.~(2006) showed that the PAH line luminosities are correlated with the IR luminosities within their starburst sample.
Schweitzer et al.~(2006) showed a similar correlation exists in a sample of QSOs due to star-formation in them.
Peeters et al.~(2004) showed a correlation between $L_{FIR}$ and $L_{\rm{PAH,}6.2}$ for a range of scales of star formation in local massive star-forming regions to normal and starburst galaxies.
The fact that the SMGs lie on this relation demonstrates the reliability of our full SED fitting to derive the total infrared luminosity and confirms that SMGs do have incredibly high infrared luminosities powered primarily by star formation.
Note that C1 does not stick out as an outlier in these plots, since we have removed the AGN component of its $L_{\rm{IR}}$.

\clearpage
\clearpage
The second thing which stands out in Fig.~\ref{fig:lir} is that there is significant scatter in this relation for both the low redshift starbursts and the high redshift SMGs. In the smaller panels the two larger colored error bars quantify the $1\sigma$ scatter seen in the starbursts and SMGs.
The scatter is similar for the SMGs and starbursts and is considerably higher than would be implied by the measurement uncertainty. 
Other properties of individual objects, such geometry of the starburst region, spatial distribution of the dust, viewing angle relative to a merging event, clumpiness of the star formation, etc. are likely playing a role in this scatter. 

We found no correlations between the IR luminosity and equivalent widths of the PAH lines in SMGs. This is consistent with what was found in Brandl et al.~(2006) for the local starburst galaxies, and shows that both the mid-IR continuum and the PAH strength scale together with increasing $L_{\rm{IR}}$.
We also find no correlation between X-ray luminosity and $L_{\rm{IR}}$ (or the PAH luminosities) for the SMGs, a correlation which would have been expected had the IR emission been dominated by an AGN.

Using the Kennicutt (1998) relation between $L_{\rm{IR}}$ and SFR, we plot the corresponding SFR on the top axis of Fig.~\ref{fig:lir}. The Kennicutt (1998) relation between $L_{\rm{IR}}$ and SFR assumes a Salpeter (1955) initial mass function and applies to starbursts with ages less than $100\,$Myr.
Assuming that both the PAH and far-IR emission are tracing the star formation, we can use the Kennicutt (1998) relation and the above PAH line luminosity to $L_{\rm{IR}}$ correlations to determine the relation between SFR and the PAH line luminosity. We use the 6.2$\,\mu$m PAH line since it is the least affected by the silicate absorption and line blending. We find that given the PAH luminosity, one can calculate the SFR using:

\begin{equation}
\label{equ:sfr}
\rm{SFR\,[M_{\odot} \rm{yr}^{-1}]} \simeq 10^{-7} \, {\it L}_{\rm{PAH,}6.2}\,[L_{\odot}].\\
\end{equation}

Similar (although perhaps slightly less accurate) relations can be found for $L_{\rm{PAH,}7.7}$ and $L_{\rm{PAH,}11.3}$ using Eq.~\ref{equ:lir}.

\section{Discussion}
\label{sec:dis}
We have shown from the mid-IR spectra of high redshift SMGs that these systems are starburst-dominated and contain only a small
contribution from the AGN. Despite, the fact that X-ray observations suggest that most SMGs harbor an AGN, only a 
small fraction of SMGs (2/13) contain an AGN which contributes significantly to the mid-IR luminosity. 
When extrapolated to the total IR luminosity, all SMGs are dominated ($>50\%$ of $L_{\rm{IR}}$) by star formation and not AGN activity.

From Table \ref{tab:lines} we see that the median $L_{\rm{IR}}^{SB}$ for our sample is $4\times10^{12}\rm{L_{\odot}}$, which corresponds to a SFR of around $700\,\rm{M_{\odot}yr^{-1}}$. This is an extremely high rate of star formation, capable of creating a massive $10^{11}\rm{M_{\odot}}$ galaxy in only $10^{8}\,$years. 
Coupled with the low number density of SMGs compared to other massive galaxies at high redshift, this leads us to suspect that SMGs are a short-lived phase in the evolution of massive galaxies. 

SMGs are only one of several populations of high redshift galaxies which are ultraluminous in the infrared. Many $z\sim2$ galaxies selected in deep 24$\,\mu$m surveys also qualify as ULIRGs (e.g.~Daddi et al.~2005; Yan et al.~2005, 2007). These other ULIRGs are generally not detected at submm wavelengths. What then are the fundamental differences between various populations of high redshift ULIRGs?

One of the main factors which could cause a bifurcation in the population of ULIRGs at high redshift is the presence or absence of a bolometrically significant AGN. The evolutionary scenario proposed by Sanders et al.~(1988) shows massive galaxies going through several main stages en route to becoming a massive elliptical. It starts with an IR luminous phase most likely triggered by a massive merger. During this stage of intense star formation, the AGN is also growing. As the AGN becomes larger it begins to feedback on the galaxy, eventually quenching the star formation completely by blowing off all the dust and gas. Thus begins the QSO phase, where the AGN is free to dominate the emission. Eventually the QSO consumes its gas 
and we end up with a quiescent massive elliptical galaxy. In this scenario, the submm emission will be at a maximum during the initial IR luminous phase (Springel, Di Matteo \& Hernquist 2005). However, one uncertainty in this scenario is the timescale. In particular, how long is the IR luminous phase and how long does it take the AGN to develop? 

Since most ULIRGs at $z\sim2$ show a mix of AGN and starburst activity (e.g.~Houck et al.~2005; Sajina et al.~2007) they fit into this evolutionary scenario at the stage where the AGN has developed but before it has had a chance to expel all the gas and dust, since the IR luminosities are high. Since SMGs show less of an AGN contribution, it is likely that they represent a slightly earlier phase of this scenario, before the AGN has time to develop fully and is significant to the IR luminosity. This is also consistent with the finding that SMGs have cool dust temperatures (Pope et al.~2006), partly because the star-formation is taking place on extended spatial scales and
partly because the AGN is not yet energetically important for heating the dust. Other high redshift ULIRGs, like those selected at 24$\,\mu$m (e.g.~Sajina et al.~2007), should have warmer dust temperatures (or have overestimated infrared luminosities), otherwise they would be detected in the submm. Therefore, they likely represent the stage after the SMG phase in this evolutionary scenario. 

For local ULIRGs, there is evidence which shows that the luminosity of CO emission decreases as the merger progresses (Rigopoulou et al.~1999), which is consistent with decreasing submm emission in the later stages of the merger. 
This idea is also consistent with morphological studies which show that SMGs are often tidally disrupted systems (Chapman et al.~2003; Conselice et al.~2003; Pope et al.~2005). Alternatively, SMGs and other high redshift ULIRGs might appear different because of different merger progenitors. 

Since the number density of SMGs is much less than that of all 24$\,\mu$m selected IR-luminous galaxies, then this phase in the evolutionary sequence must be shorter. 
We can make a rough estimate for the duration of the submm luminous phase. In the 2$\,$Gyrs between $1.5<z<3$ there are roughly 20 SMGs and 500 24$\,\mu$m-selected galaxies (with $L_{\rm{IR}}>10^{11}\rm{L_{\odot}}$, see Chary et al.~2004) within a portion of the GOODS-N field. Assuming that all IR-luminous 24$\,\mu$m-selected galaxies will undergo a submm luminous stage during this period of time, we calculate the duration of the submm luminous phase to be on the order of $10^{8}\,$yrs\footnote{It is not clear what fraction of 24$\,\mu$m-selected galaxies will undergo a submm-luminous phase since SMGs are known to be massive and could represent only the most extreme galaxies. Nevertheless, this calculation provides a useful reality check on the expected duration of the SMG phase.}.
As mentioned above, this timescale is what is needed to create a galaxy with a stellar mass of $\sim10^{11}\rm{M_{\odot}}$ (Borys et al.~2005) at a rate of $\sim1000\,\rm{M_{\odot}yr^{-1}}$. This timescale is consistent with those derived from gas masses of SMGs (Greve et al.~2005).

\section{Conclusions}

{\it Spitzer} IRS spectroscopy has been obtained for a sample of 13 SMGs brighter than 200$\,\mu$Jy at 24$\,\mu$m. This has effectively confirmed the identification of these SMG counterparts, and in some cases has provided a new redshift estimate. 
The SMGs show strong PAH emission and, on average, only a small continuum contribution from hot dust.

We have explored several diagnostics from the mid-IR spectra of SMGs to determine the level of AGN contributing to the luminosity at these wavelengths. All of them seem to converge on a picture in which SMGs are starburst-dominated systems with at most a 30\% contribution from an AGN at mid-IR wavelengths. Their mid-IR spectra are similar to a scaled spectrum of local starburst galaxies like M82. 

The classification of an SMG as AGN or starburst-dominated from the mid-IR often disagrees with the classification from the X-rays. While X-ray observations are better suited for determining the presence of an AGN in an SMG, the mid-IR spectra can determine how important the AGN is to the total infrared luminosity since it is directly detecting the hot dust emission and not subject to details about the geometry
of the obscuring material.

Full IR SED fits to the mid-IR data, far-infrared and radio photometry show that SMGs are best-fit by scaled up versions of local IR-luminous galaxy templates with additional extinction from the Draine (2003) extinction curves. These models have cool dust temperatures ($T=32\,$K) and high $L_{\rm{IR}}$s which imply SFRs of $\sim700\,\rm{M_{\odot}yr^{-1}}$ assuming a Salpeter initial mass function. 

SMGs lie on the local relation between $L_{\rm{IR}}$ and $L_{\rm{PAH,}6.2}$ (or $L_{\rm{PAH,}7.7}$ or $L_{\rm{PAH,}11.3}$), which means that the PAH line flux can be used to estimate the SFRs in these systems, albeit with large uncertainties. Equivalent widths, which have been the focus of some other studies, are much less useful in this regard. 

SMGs appear to be an earlier phase in the evolution of massive galaxies than other high redshift ULIRGs. The average high redshift ULIRG
shows warmer dust and less PAH emission. It is therefore likely that the AGN in SMGs has not yet grown strong enough to
heat the dust substantially and destroy the PAH molecules.

We have been able to put constraints on the contribution from AGN activity to the restframe 5--12$\,\mu$m mid-IR emission, and have extrapolated this to far-IR wavelengths.
Without more data from 30--100$\,\mu$m rest frame it is difficult to directly probe the emission at far-IR wavelengths.
We will be able to model this part of the spectrum in more detail once data from the {\it Herschel Space Observatory}\footnote{\tt{http://herschel.esac.esa.int/}} and SCUBA-2 (Holland et al.~2006) become available.
Details of the SMG phase will be understood only after the spatially resolved spectroscopy capabilities of ALMA\footnote{\tt{http://www.eso.org/projects/alma/}} become available.

\acknowledgments
We would like to thank the referee for his/her helpful suggestions which improved the quality of this paper. 
We thank Anna Sajina for helpful discussions and for providing composite IRS spectra of their high redshift {\it Spitzer} 24\,$\mu$m-selected ULIRGs.
We are very grateful to Bernhard Brandl for providing the IRS spectra of local starburst galaxies and Benjamin Magnelli for providing information from the completeness simulations of the 70$\,\mu$m images.
This work was supported by the Natural Sciences and Engineering Research Council of Canada and the Canadian Space Agency. DMA acknowledges the Royal Society for support.
This work is based on observations made with the Spitzer Space Telescope, which is operated by the Jet Propulsion Laboratory, California Institute of Technology under a contract with NASA. Support for this work was provided by NASA through an award issued by JPL/Caltech.
The IRS was a collaborative venture between Cornell University and Ball Aerospace Corporation funded by NASA through the Jet Propulsion Laboratory and Ames Research Center.

\appendix

\section{{\it Spitzer} IRS noise}
\label{app:rms}

\begin{figure}
\plotone{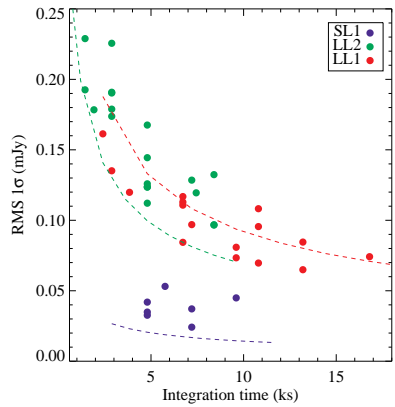}
\caption{Observed $1\sigma$ RMS as a function of integration time for our IRS staring mode observations. All observations from our {\it Spitzer} GO2-20456 program are shown here. The dashed curves show the predicted values from SPEC-PET for each order. 
We confirm that the sensitivity goes as $\sigma\propto t^{-1/2}$ even for very long integrations.
The scatter in the measure values is probably dominated by the accuracy of the sky subtraction in these crowded fields. 
Given the scatter, our observed sensitivities are consistent with those from SPEC-PET. 
\label{fig:rms}}
\end{figure}

Fig.~\ref{fig:rms} shows the final RMS of our spectra as a function of the integration time and spectral order. Values from the {\it Spitzer} Science Center sensitivity calculator; SPEC-PET\footnote{\tt{http://ssc.Spitzer.caltech.edu/tools/specpet/}} are given for comparison as the dashed curves. We confirm that the sensitivity goes as $\sigma (\rm{mJy})=a\times t (\rm{s})^{-1/2}$ even for very long integrations.
The scatter in the measured values is probably dominated by the accuracy of the sky subtraction in these crowded fields. 
Given the scatter, our observed sensitivities are consistent with those from SPEC-PET. 
The decrease in sensitivity in the SL1 order could be related to our choice of 240 second ramps in fields with backgrounds $>20\,$MJy$\,$sr$^{-1}$ at 24$\,\mu$m, which results in incorrect droop correction (see the {\it Spitzer} Observer's Manual).

\clearpage

\section{Notes on individual sources}
\label{ap:sources}
Here we describe several of our sources in more detail. Sources which are not discussed here are relatively straightforward in terms of interpreting their mid-IR spectra. 

\subsection{GN06}
The optical and IRS redshifts disagree for this SMG. The IRS redshift is based on 3 PAH features: 6.2, 7.7 and 11.3$\,\mu$m and the optical spectroscopic redshift of 1.865 is inconsistent with the positions of these emission features, even within the errors. The optical spectroscopic redshift of GN06 comes from Chapman et al.~(2005) where they note that the optical galaxy was offset from the radio counterpart by about an arcsecond. From the images of GN06 in Fig.~\ref{fig:post}, we see there are two optical knots both of which are offset from the counterpart which is detected in the radio and mid-IR. For this source, we conclude that the optical and mid-IR spectra are picking up different galaxies and the submm emission is coming from the mid-IR detected galaxy.

\subsection{C2}
C2 has the largest discrepancy between the IRS and UV/optical spectroscopic redshifts. The IRS redshift for this source is based on  the 7.7 and 11.3$\,\mu$m PAH features and there is also a hint of the 8.6$\,\mu$m line. The optical spectroscopic redshift would have placed the 7.7$\,\mu$m PAH line right in the middle of this spectrum and therefore it is inconsistent with the IRS spectrum. The optical spectroscopic redshift comes from detections of both H$\alpha$ and Lyman$\alpha$ emission (Swinbank et al.~2004; Chapman et al. 2005). The images in Fig.~\ref{fig:post} show an extended optical galaxy. This is a curious source, however since the mid-IR is more likely to trace the submm emission, we conclude that the IRS redshift is correct for this SMG.

\subsection{C1}
This SMG is not detected in the deep {\it Chandra} 2 Msec imaging of GOODS-N. Furthermore, the optical spectrum from Swinbank et al.~(2004) clearly shows H$\alpha$, NII and the SII doublet and the low NII to H$\alpha$ ratio is used to classify this SMG as a starburst.
However the mid-IR spectrum of C1 clearly shows a steeply rising continuum indicative of an AGN emission. This is also apparent from the extreme {\it Spitzer} colors shown in Fig.~\ref{fig:cc}.
The mid-IR spectrum also shows PAH emission and the ratio of PAH luminosity to IR luminosity for C1 is significantly lower than those seen in local starbursts. 
We conclude that this SMG harbors an Compton-thick AGN which accounts for 44\% of the total IR luminosity. This is the only SMG which contains a bolometrically significant AGN. 
Interestingly, C1 has not been detected in CO (Greve et al.~2005).
From the X-ray flux limit, we can place a limit on the column density of $N_{\rm{H}}\gg 10^{24}\rm{cm}^{-2}$.
C1 is relatively bright at 24$\,\mu$m ($1.2\,$mJy) suggesting that these types of obscured AGN at high redshift should be detectable in even shallow {\it Spitzer} surveys.

\subsection{GN39 double system}

One further SMG deserves special discussion and that is GN39.
This SMG has two radio/MIPS counterparts separated by 8 arcseconds, which we refer to as GN39a and GN39b.
Both are confirmed to lie at $z=1.996$ (Swinbank et al.~2004; Chapman et al.~2005), therefore it is likely that they are both contributing to the submm emission. The separation corresponds to 70 kpc at this redshift.

Our IRS observations picked up both counterparts and, with a narrow extraction window, we were able to extract a separate spectrum of each component. 
Some of the flux from GN39a falls into the spectral
extraction window of GN39b however a comparison of the 24$\,\mu$m flux from the IRS spectra to that from the MIPS 24$\,\mu$m image confirms that only 1/3 of the flux in the IRS spectrum of GN39b is from GN39a. 
The extracted spectra confirm that these two sources are indeed at the same redshift and both spectra look very similar to each other and to the rest of the SMGs, supporting the idea that they are both submm emitters. Furthermore, both components are X-ray detected and classified as obscured AGN by Alexander et al.~(2005), although the two IRS spectra show a completely starburst-dominated system at mid-IR wavelengths. 

Since these two SMGs are at the same redshift, we know that they must be associated. 
Given the physical distance between them, larger than those seen in most local ULIRGs (Murphy et al.~1996),
we could be seeing the early stage in the
merger of two massive galaxies (Springel, Di Matteo \& Hernquist 2005). 
Note that GN04, GN07 and GN19 also have double radio counterparts, which are thought to be at the same redshift. However, in these sources, the two components are separated by only 3 arcseconds and therefore they do not resolve into two separate 24$\,\mu$m sources. Thus the IRS spectra for these sources contain contributions from both components.

\clearpage

%
%

\begin{deluxetable}{ccrrrrrr}
\tabletypesize{\scriptsize}
\tablecaption{Coordinates and flux densities of IRS targets}
\tablehead{
\colhead{SMG Name} & \colhead{SMG ID} &
\colhead{RA} & \colhead{DEC}  &
 \colhead{$S_{16}$} &  \colhead{$S_{24}$} & \colhead{$S_{70}$}  \\ 
 \colhead{SMMJ..} & &  \colhead{J2000} & \colhead{J2000}  & \colhead{(mJy)} & \colhead{(mJy)} &  \colhead{(mJy)}  
\label{tab:targets}
}
\startdata 
123653.1+621120 &  GN31   & 12:36:53.22  & 62:11:16.7  &  0.301 $\pm$ 0.007 &  0.367 $\pm$ 0.006  & $<3.5$   \\ 
123635.5+621238  &  GN26  &  12:36:34.51 &  62:12:40.9   & 0.992 $\pm$ 0.012  & 0.446 $\pm$ 0.005 & $13.9 \pm 1.8$  \\ 
123701.2+621147  &  GN17  & 12:37:01.59 & 62:11:46.2 & 0.209 $\pm$ 0.006 & 0.710 $\pm$ 0.008 & $<3.5$  \\ 
123622.6+621629 &   C2  & 12:36:22.66 &  62:16:29.5  &  0.098 $\pm$ 0.006  & 0.414 $\pm$ 0.007  &   $<3.5$  \\ 
123555.1+620901  &  C3  &  12:35:55.13 &  62:09:01.6 & n/a &  0.374 $\pm$ 0.009 & $<3.5$   \\ 
123711.1+621325  &  GN39a  & 12:37:11.37 &  62:13:31.1  & 0.097 $\pm$ 0.005   & 0.537 $\pm$ 0.009  & $<3.5$   \\ 
123711.9+621331   &   GN39b   & 12:37:11.97  & 62:13:25.8 &  0.051 $\pm$ 0.005 &  0.225 $\pm$ 0.007 &  $<3.5$  \\
123621.3+621711  &  GN07  &  12:36:21.27 &  62:17:08.1  & 0.048 $\pm$ 0.007 &   0.370 $\pm$ 0.011  & $<3.5$  \\ 
123618.7+621553  &  GN06   & 12:36:18.33  & 62:15:50.4  &  0.036 $\pm$ 0.006  & 0.330 $\pm$ 0.008 & $<3.5$  \\ 
123600.2+621047 &  C1   & 12:36:00.16 & 62:10:47.3 & 0.478 $\pm$ 0.011 &1.220 $\pm$ 0.008  & $5.5 \pm 1.2$  \\ 
123618.8+621008  &  GN05   & 12:36:19.13  & 62:10:04.3  &  0.044 $\pm$ 0.007  & 0.215 $\pm$ 0.006 & $<3.5$   \\ 
123707.7+621411  &  GN19   & 12:37:07.19  & 62:14:08.0  & $<0.02$  & 0.280 $\pm$ 0.015 & $<3.5$   \\ 
123616.6+621520  &  GN04   & 12:36:16.11 &  62:15:13.5  &  0.060 $\pm$ 0.006 &  0.303 $\pm$ 0.007  & $<3.5$  \\ 
\enddata
\end{deluxetable}

\clearpage

\begin{deluxetable}{crrrcr}
\tablecaption{{\it Spitzer} IRS observations.}
\tablehead{
\colhead{SMG ID} &
 \multicolumn{4}{c}{Integration time}\\ 
& \colhead{SL1\,($\times240$s)} & \colhead{LL2\,($\times120$s)}   & \colhead{LL1\,($\times120$s)}   & \colhead{Total\,(hr)}
\label{tab:obs}
}
\startdata 
GN31   &    & 35$\times$2 & & 2.3 \\ 
GN26  &   15$\times$2  & 20$\times$2 & [12$\times$2]  & 3.3 [4.1] \\ 
GN17  & 20$\times$2 & 12$\times$2 &  & 3.5 \\ 
C2  &    &  & 45$\times$2  & 3.0 \\ 
C3  &  & 31$\times$2 & 40$\times$2  & 4.7 \\ 
GN39a  &  & 20$\times$2 & 28$\times$2  & 3.2 \\ 
GN39b   &     & [20$\times$2]$^{a}$  & [28$\times$2] & [3.2]\\
GN07  & & 35$\times$2 & 45$\times$2  &  5.3 \\ 
GN06   &    & 35$\times$2 & 55$\times$2  & 6.0 \\ 
C1   &  & 6$\times$2  & 10$\times$2   & 1.1 \\ 
GN05   & & & 55$\times$2 & 3.7\\ 
GN19   & & [20$\times$2] & [28$\times$2] & [3.2] \\ 
GN04   &  & & 70$\times$2 & 4.7\\ 
\enddata
\tablenotetext{a}{Sources with time in square brackets are secondary sources observed in the slits of the primary targets. }
\end{deluxetable}

\clearpage

\begin{deluxetable}{crrrc}
\tablecaption{Measured IRS redshifts compared to optical spectroscopic and photometric redshifts. }
\tablehead{
 \colhead{SMG ID} & 
 \colhead{$i$ mag (AB)$^{a}$}  &
 \colhead{$z_{\rm{optical}}$$^{d}$} &  
 \colhead{$z_{\rm{IRS}} \pm 1\sigma$} &
\colhead{Comment}
\label{tab:redshift}
}
\startdata
GN31 & 21.8  & 0.935$^{h}$ & 0.93 $\pm$ 0.03  &   \\
GN26 & 22.7  &  1.219$^{e}$  & 1.23 $\pm$ 0.01 &  \\
GN17 & 27.7  & 1.72$^{g}$ & 1.73 $\pm$ 0.01  &  new IRS redshift \\
C2 & 25.5  & 2.466$^{e,f}$ & 1.79 $\pm$ 0.04  & inconsistent \\
C3 &  23.5$^{b}$ & 1.875$^{e}$ & 1.88 $\pm$ 0.02 &  \\
GN39a & 25.9  & 1.996$^{f}$ & 1.98 $\pm$ 0.01 &  \\
GN39b & 26.4  & 1.992$^{e}$ & 1.99 $\pm$ 0.04 &  \\
GN07 & 27.8/23.9$^{c}$  & 1.988$^{e}$, 1.992$^{f}$ & 1.99 $\pm$ 0.02 &  \\
GN06 & 27.8  & 1.865$^{e}$ & 2.00 $\pm$ 0.03  &  inconsistent \\
C1 &  24.8 & 1.994$^{e}$, 2.002$^{f}$ & 2.01 $\pm$ 0.05 & \\
GN05 & 24.9  & 2.60$^{g}$ & 2.21 $\pm$ 0.03 & new  IRS redshift\\
GN19 & 25.4/$>$28$^{c}$  & 2.484$^{e}$, 2.490$^{f}$ & 2.48 $\pm$ 0.03  &   \\
GN04 & 26.2/$>$28$^{c}$  & 2.578$^{e}$ & 2.55 $\pm$ 0.01  &  \\
\enddata
\tablenotetext{a}{$i$ magnitudes are from the GOODS ACS observations unless otherwise noted. }
\tablenotetext{b}{Capak et al.~(2004). }
\tablenotetext{c}{These SMGs have two counterparts within the IRS aperture. }
\tablenotetext{d}{Optical redshifts are spectroscopic unless otherwise noted. }
\tablenotetext{e}{Chapman et al.~(2005). }
\tablenotetext{f}{Swinbank et al.~(2004). }
\tablenotetext{g}{Pope et al.~(2006), photometric redshift. }
\tablenotetext{h}{Cowie et al.~(2004). }
\end{deluxetable}

\clearpage

\begin{deluxetable}{ccccr}
\tabletypesize{\scriptsize}
\tablecaption{Classification of mid-IR spectra compared with classifications based on X-ray imaging and UV/optical spectroscopy.
}
\tablehead{
 \colhead{SMG ID} & \multicolumn{3}{c}{Classification$^{a}$} &
 Mid-IR \\
& X-ray$^{b}$ & UV/Op$^{c}$ & Mid-IR & continuum\%$^{d}$
\label{tab:agn}
}
\startdata
GN31 & U & n/a & SB &  $<45$ \\
GN26 & SB & SB  & SB  & $<15$ \\
GN17 & SB & n/a & SB &  $<29$ \\
C2 &  AGN & SB & SB &  $<34$   \\
C3 & AGN &  SB & SB+AGN & 48 \\
GN39a & oAGN & SB & SB & $<35$ \\
GN39b & oAGN & SB & SB & $<10$   \\
GN07 & SB & SB &SB+weak AGN  & 18\\
GN06 & SB  & SB & SB+AGN &  47  \\
C1 & SB  & SB & AGN+weak SB  & 82 \\
GN05 & U & n/a & SB & $<7$   \\
GN19 & oAGN & SB  & SB  & $<14$  \\
GN04 & AGN & SB & AGN+SB  & 61 \\
\enddata
\tablenotetext{a}{oAGN = obscured AGN, SB=starburst, U=undetected.}
\tablenotetext{b}{Classified as AGN if detected in the hard X-ray band.}
\tablenotetext{c}{Classification from Chapman et al.~(2005) and/or Swinbank et al.~(2004).}
\tablenotetext{d}{Percentage of mid-IR flux that comes from continuum emission.}
\end{deluxetable}

\clearpage

\begin{deluxetable}{crrrrrrrrr}
\tabletypesize{\scriptsize}
\tablecaption{PAH strengths and infrared luminosities. 
}
\tablehead{
 \colhead{SMG ID} & 
\multicolumn{3}{c}{PAH luminosity ($10^{9}\rm{L_{\odot}}$)} & 
\multicolumn{3}{c}{PAH equivalent width$^{a}$ ($\mu$m)} & 
\multicolumn{3}{c}{$L_{\rm{IR}}$ ($10^{12}$L$_{\odot}$)} \\
& 6.2$\,\mu$m & 7.7$\,\mu$m  & 11.3$\,\mu$m & 6.2$\,\mu$m & 7.7$\,\mu$m  & 11.3$\,\mu$m   & SB & AGN &Total 
\label{tab:lines}
}
\startdata
GN31 & ...   &  $3.4 \pm 1.2$  &    ...  &   ...     &   $0.32 \pm 0.11$ & ...   &   $>0.17$ & $<0.10$ & 0.27 \\ 
GN26 & $12.2 \pm 1.2$   &     $42.9 \pm 5.1$   &     $8.1 \pm 1.4$    &  $0.38 \pm 0.04$     &  $1.00 \pm 0.13$     &  $1.18 \pm 0.54$ &  $>3.5$ & $<0.1$ & 3.6 \\ 
GN17 & $13.0 \pm 4.1$ & ... & ... & $0.22 \pm 0.07$ & ...  & ... &  $>1.3$ & $<0.7$ & 2.0\\
C2 &  ...     &  $31.1 \pm 6.3$    &   $7.0 \pm 1.8$      & ... & $1.08 \pm 0.30$    &    $1.00 \pm 0.29$ & $>2.8$ & $<0.5$  & 3.3\\
C3 & $<19.2$  & $ 9.2 \pm 2.7$  &     $7.9 \pm 2.7$   & $<0.50$ &    $0.25 \pm 0.08$   &  $0.45 \pm 0.16$ & 5.2 & 0.8 & 6.0\\ 
GN39a &  $13.9 \pm 4.9$ &  $39.2 \pm 12.9$  & $<36.3$ &   $0.66 \pm 0.26$ & $0.86 \pm 0.30$  &  $<0.86$  & $>4.2$ & $<0.8$ & 5.0\\
GN39b &  $12.3 \pm 5.6$    &     $23.1 \pm 5.1$      & $<29.4$ &    $1.09 \pm 0.65$     &  $0.59 \pm 0.14$       & $<1.14$ &  $>2.7$ & $<0.1$ & 2.8\\
GN07 & $12.5 \pm 2.6$   &   $17.6 \pm 3.2$ &      $9.9 \pm 1.8$  &    $1.10 \pm 0.38$  &   $0.41 \pm 0.08$  &    $1.58 \pm 0.47$  & 6.7 &  0.2 & 6.9 \\
GN06 &  $18.0 \pm 1.5$  &   $26.3 \pm 4.4$   &   $10.5 \pm 2.2$   &   $0.83 \pm 0.10$   &   $0.93 \pm 0.16 $   &   $2.95 \pm 1.38$ & 5.7 &  0.7 & 6.4\\ 
C1 & $6.3 \pm 1.5$   &    $59.1 \pm 12.6$      &     $<18.4$      &    $0.05\pm0.01$   &  $0.54 \pm 0.14$  &  $<0.18$      & 5.4 & 4.2 & 9.6 \\
GN05 & ... & $12.9 \pm 4.3$ & ...  & ... & $0.44 \pm 0.15$ & ... & $>3.9$ & $<0.1$ & 4.0\\
GN19 & $12.5 \pm 3.0$  &   $32.5 \pm 9.0$          & ... & $0.59 \pm 0.20$   &    $0.77 \pm 0.22$      & ... & $>11.9$ & $<0.4$ & 12.3\\ 
GN04 & $<23.7$  & $43.3 \pm 9.9$ & ... & $<0.45$ & $0.63 \pm 0.15$ & ... & 4.5 & 2.1 & 6.6 \\ 
\enddata
\tablenotetext{a}{The equivalent widths are all in rest frame. }
\end{deluxetable}


\begin{thebibliography}{}


\bibitem[\protect\citeauthoryear{Alexander et al.}{2003b}]{Alexander03} Alexander D.M., et al., 2003, AJ, 126, 539 
\bibitem[\protect\citeauthoryear{Alexander et al.}{2005}]{Alexander05} Alexander D.M., Bauer F.E., Chapman S.C., Smail I., Blain A.W., Brandt W.N., Ivison R.J., 2005, ApJ, 632, 736
\bibitem[Allamandola et al.(1999)]{1999ApJ...511L.115A} Allamandola, L.~J., Hudgins, D.~M., \& Sandford, S.~A.\ 1999, \apjl, 511, L115 
\bibitem[Aretxaga et al.(2007)]{2007astro.ph..2503A} Aretxaga, I., et al.\ 2007, MNRAS, 379, 1571
\bibitem[Armus et al.(2007)]{2007ApJ...656..148A} Armus, L., et al.\ 2007, \apj, 656, 148
\bibitem[Bauer et al.(2002)]{2002AJ....124.2351B} Bauer, F.~E., Alexander, D.~M., Brandt, W.~N., Hornschemeier, A.~E., Vignali, C., Garmire, G.~P., \& Schneider, D.~P.\ 2002, \aj, 124, 2351 
\bibitem[\protect\citeauthoryear{Blain et al.}{2002}]{Blain02} Blain A.W., Smail I., Ivison R.J., Kneib J.-P., Frayer D.T., 2002, PhR, 369, 111 
\bibitem[\protect\citeauthoryear{Borys et al.}{2003}]{PaperI} Borys C., Chapman S., Halpern M., Scott D., 2003, MNRAS, 344, 385 
\bibitem[\protect\citeauthoryear{Borys et al.}{2005}]{2005ApJ...635..853B} Borys C., Smail I., Chapman S.C., Blain A.W., Alexander D.M., Ivison R.J., 2005, ApJ, 635, 853 
\bibitem[Brandl et al.(2006)]{2006ApJ...653.1129B} Brandl, B.~R., et al.\ 2006, \apj, 653, 1129
\bibitem[Brandt \& Hasinger(2005)]{2005ARA&A..43..827B} Brandt, W.~N., \& Hasinger, G.\ 2005, \araa, 43, 827 
\bibitem[\protect\citeauthoryear{Capak et al.}{2004}]{Capak04} Capak P., et al., 2004, AJ, 127, 180
\bibitem[Chapman et al.(2003)]{2003ApJ...599...92C} Chapman, S.~C., Windhorst, R., Odewahn, S., Yan, H., \& Conselice, C.\ 2003, \apj, 599, 92 
\bibitem[\protect\citeauthoryear{Chapman et al.}{2005}]{2005ApJ...622..772C} Chapman S.~C., Blain A.~W., Smail I., Ivison R.~J., 2005, ApJ, 622, 772
\bibitem[Charmandaris et al.(1999)]{1999Ap&SS.266...99C} Charmandaris V., et al., 1999, \apss, 266, 99 
\bibitem[\protect\citeauthoryear{Chary \& Elbaz}{2001}]{CE01} Chary R., Elbaz D., 2001, ApJ, 556, 56 (CE01)
\bibitem[Chary et al.(2004)]{2004ApJS..154...80C} Chary, R., et al.\ 2004, \apjs, 154, 80 
\bibitem[Coppin et al.(2006)]{2006MNRAS.372.1621C} Coppin K., et al., 2006, \mnras, 372, 1621 
\bibitem[Conselice et al.(2003)]{2003ApJ...596L...5C} Conselice, C.~J., Chapman, S.~C., \& Windhorst, R.~A.\ 2003, \apjl, 596, L5 
\bibitem[\protect\citeauthoryear{Cowie et al.}{2004}]{Cowie04} Cowie L.L., Barger A.J., Hu E.M., Capak P., Songaila A., 2004, AJ, 127, 3137 
\bibitem[\protect\citeauthoryear{Daddi et al.}{2005}]{2005ApJ...631L..13D} Daddi E., et al., 2005, ApJ, 631, L13
\bibitem[Daddi et al.(2007)]{2007arXiv0705.2832D} Daddi E., et al., 2007, ApJ in press, (astro-ph/0705.2832) 
\bibitem[Desai et al.(2007)]{2007arXiv0707.4190D} Desai V., et al., 2007, ApJ, 669, 810 
\bibitem[Dickinson et al.(2003)]{2003mglh.conf..324D} Dickinson, M., Giavalisco, M., \& The Goods Team 2003, The Mass of Galaxies at Low and High Redshift, 324 
\bibitem[Draine(2003)]{2003ARA&A..41..241D} Draine, B.~T.\ 2003, \araa, 41, 241
\bibitem[Draine \& Li(2007)]{2007ApJ...657..810D} Draine, B.~T., \& Li, A.\ 2007, \apj, 657, 810 
\bibitem[\protect\citeauthoryear{Dunne et al.}{2000}]{2000MNRAS.315..115D} Dunne L., Eales S., Edmunds M., Ivison R., Alexander P., Clements D.L., 2000, MNRAS, 315, 115 
\bibitem[Farrah et al.(2007)]{2007ApJ...667..149F} Farrah, D., et al.\ 2007, \apj, 667, 149 
\bibitem[Fischer et al.(1999)]{1999Ap&SS.266...91F} Fischer J., et al., 1999, \apss, 266, 91 
\bibitem[F{\"o}rster Schreiber et al.(2003)]{2003A&A...399..833F} F{\"o}rster Schreiber, N.~M., et al., 2003, \aap, 399, 833
\bibitem[Frayer et al.(2006)]{2006ApJ...647L...9F} Frayer, D.~T., et al.\ 2006, \apjl, 647, L9 
\bibitem[Genzel et al.(1998)]{1998ApJ...498..579G} Genzel, R., et al.\ 1998, \apj, 498, 579 
\bibitem[\protect\citeauthoryear{Giavalisco et al.}{2004}]{Giavalisco04} Giavalisco M., et al., 2004, ApJ, 600, L93 
\bibitem[\protect\citeauthoryear{Greve et al.}{2005}]{2005MNRAS.359.1165G} Greve T.R., et al., 2005, MNRAS, 359, 1165
\bibitem[Hauser \& Dwek(2001)]{2001ARA&A..39..249H} Hauser, M.~G., \& Dwek, E.\ 2001, \araa, 39, 249 
\bibitem[Higdon et al.(2004)]{2004PASP..116..975H} Higdon, S.~J.~U., et al.\ 2004, \pasp, 116, 975 
\bibitem[\protect\citeauthoryear{Holland et al.}{1999}]{Holland99} Holland W.S., et al., 1999, MNRAS, 303, 659 
\bibitem[\protect\citeauthoryear{Holland et al.}{2006}]{Holland06} Holland W.S., et al., 2006, SPIE conference, 6275, 45
\bibitem[Houck et al.(2004)]{2004ApJS..154...18H} Houck, J.~R., et al.\ 2004, \apjs, 154, 18 
\bibitem[Houck et al.(2005)]{2005ApJ...622L.105H} Houck, J.~R., et al.\ 2005, \apjl, 622, L105 
\bibitem[Huynh et al.(2007)]{2007ApJ...659..305H} Huynh, M.~T., Pope, A., Frayer, D.~T., \& Scott, D.\ 2007, \apj, 659, 305 
\bibitem[\protect\citeauthoryear{Ivison et al.}{2002}]{Ivison02} Ivison R.J., et al., 2002, MNRAS, 337, 1 
\bibitem[\protect\citeauthoryear{Ivison et al.}{2004}]{Ivison04} Ivison R.J., et al., 2004, ApJS, 154, 124
\bibitem[\protect\citeauthoryear{Kennicutt}{1998}]{1998ARA&A..36..189K} Kennicutt R.~C., 1998, ARA\&A, 36, 189 
\bibitem[Kov{\'a}cs et al.(2006)]{2006ApJ...650..592K} Kov{\'a}cs, A., Chapman, S.~C., Dowell, C.~D., Blain, A.~W., Ivison, R.~J., Smail, I., \& Phillips, T.~G.\ 2006, \apj, 650, 592 
\bibitem[Lagache et al.(2005)]{2005ARA&A..43..727L} Lagache, G., Puget, J.-L., \& Dole, H.\ 2005, \araa, 43, 727
\bibitem[Le Floc'h et al.(2005)]{2005ApJ...632..169L} Le Floc'h, E., et al.\ 2005, \apj, 632, 169 
\bibitem[\protect\citeauthoryear{Lilly et al.}{1999}]{Lilly99} Lilly S.J., Eales S.A., Gear W.K.P., Hammer F., Le F{\` e}vre O., Crampton D., Bond J.R., Dunne L., 1999, ApJ, 518, 641 
\bibitem[Lutz et al.(1998)]{1998ApJ...505L.103L} Lutz, D., Spoon, H.~W.~W., Rigopoulou, D., Moorwood, A.~F.~M., \& Genzel, R.\ 1998, \apjl, 505, L103 
\bibitem[Lutz et al.(1999)]{1999Ap&SS.266...85L} Lutz, D., Genzel, R., Kunze, D., Rigopoulou, D., Spoon, H.~W.~W., Sturm, E., Tran, D., \& Moorwood, A.~F.~M.\ 1999, \apss, 266, 85 
\bibitem[Lutz et al.(2004)]{2004A&A...418..465L} Lutz, D., Maiolino, R., Spoon, H.~W.~W., \& Moorwood, A.~F.~M.\ 2004, \aap, 418, 465 
\bibitem[\protect\citeauthoryear{Lutz et al.}{2005}]{Lutz05} Lutz D., Valiante E., Sturm E., Genzel R., Tacconi L.~J., Lehnert M.~D., Sternberg A., Baker A.~J., 2005a, ApJ, 625, L83 
\bibitem[Lutz et al.(2005)]{2005ApJ...632L..13L} Lutz, D., Yan, L., Armus, L., Helou, G., Tacconi, L.~J., Genzel, R., \& Baker, A.~J.\ 2005b, \apjl, 632, L13 
\bibitem[Men{\'e}ndez-Delmestre et al.(2007)]{2007ApJ...655L..65M} Men{\'e}ndez-Delmestre, K., et al.\ 2007, \apjl, 655, L65
\bibitem[Murphy et al.(1996)]{1996AJ....111.1025M} Murphy, T.~W., Jr., Armus, L., Matthews, K., Soifer, B.~T., Mazzarella, J.~M., Shupe, D.~L., Strauss, M.~A., \& Neugebauer, G.\ 1996, \aj, 111, 1025 
\bibitem[P{\'e}rez-Gonz{\'a}lez et al.(2005)]{2005ApJ...630...82P} P{\'e}rez-Gonz{\'a}lez, P.~G., et al.\ 2005, \apj, 630, 82 
\bibitem[\protect\citeauthoryear{Pope et al.}{2005}]{Pope05} Pope A., Borys C., Scott D., Conselice C., Dickinson M., Mobasher B., 2005, MNRAS, 358, 149
\bibitem[Pope et al.(2006)]{2006MNRAS.370.1185P} Pope, A., et al.\ 2006, \mnras, 370, 1185
\bibitem[\protect\citeauthoryear{Pope}{2007}]{Pope07} Pope A., 2007, PhD thesis, University of British Columbia
\bibitem[Puget \& Leger(1989)]{1989ARA&A..27..161P} Puget, J.~L., \& Leger, A.\ 1989, \araa, 27, 161 
\bibitem[Puget et al.(1996)]{1996A&A...308L...5P} Puget, J.-L., Abergel, A., Bernard, J.-P., Boulanger, F., Burton, W.~B., Desert, F.-X., \& Hartmann, D.\ 1996, \aap, 308, L5 
\bibitem[\protect\citeauthoryear{Rigopoulou et al.}{1999}]{1999AJ....118.2625R} Rigopoulou D., Spoon H.~W.~W., Genzel R., Lutz D., Moorwood A.~F.~M., Tran Q.~D., 1999, AJ, 118, 2625 
\bibitem[Sajina et al.(2005)]{2005ApJ...621..256S} Sajina, A., Lacy, M., \& Scott, D.\ 2005, \apj, 621, 256 
\bibitem[Sajina et al.(2007)]{2007arXiv0704.1765S} Sajina, A., Yan, L., Armus, L., Choi, P., Fadda, D., Helou, G., \& Spoon, H.\ 2007, ApJ, 664, 713
\bibitem[\protect\citeauthoryear{Salpeter}{1955}]{1955ApJ...121..161S} Salpeter E.E., 1955, ApJ, 121, 161 
\bibitem[Sanders et al.(1988)]{1988ApJ...325...74S} Sanders, D.~B., Soifer, B.~T., Elias, J.~H., Madore, B.~F., Matthews, K., Neugebauer, G., \& Scoville, N.~Z.\ 1988, \apj, 325, 74 
\bibitem[Sanders \& Mirabel(1996)]{1996ARA&A..34..749S} Sanders, D.~B., \& Mirabel, I.~F.\ 1996, \araa, 34, 749
\bibitem[Schweitzer et al.(2006)]{2006ApJ...649...79S} Schweitzer, M., et al.\ 2006, \apj, 649, 79  
\bibitem[Smith et al.(2007)]{2007ApJ...656..770S} Smith, J.~D.~T., et al.\ 2007, \apj, 656, 770
\bibitem[Soifer et al.(1987)]{1987ApJ...320..238S} Soifer, B.~T., Sanders, D.~B., Madore, B.~F., Neugebauer, G., Danielson, G.~E., Elias, J.~H., Lonsdale, C.~J., \& Rice, W.~L.\ 1987, \apj, 320, 238 
\bibitem[Spoon et al.(2007)]{2007ApJ...654L..49S} Spoon, H.~W.~W., Marshall, J.~A., Houck, J.~R., Elitzur, M., Hao, L., Armus, L., Brandl, B.~R., \& Charmandaris, V.\ 2007, \apjl, 654, L49 
\bibitem[Springel et al.(2005)]{2005MNRAS.361..776S} Springel, V., Di Matteo, T., \& Hernquist, L.\ 2005, \mnras, 361, 776 
\bibitem[Sturm et al.(2000)]{2000A&A...358..481S} Sturm, E., Lutz, D., Tran, D., Feuchtgruber, H., Genzel, R., Kunze, D., Moorwood, A.~F.~M., \& Thornley, M.~D.\ 2000, \aap, 358, 481 
\bibitem[\protect\citeauthoryear{Swinbank et al.}{2004}]{2004ApJ...617...64S} Swinbank A.M., Smail I., Chapman S.C., Blain A.W., Ivison R.J., Keel W.C., 2004, ApJ, 617, 64 
\bibitem[Tacconi et al.(2006)]{2006ApJ...640..228T} Tacconi L.J., et al., 2006, \apj, 640, 228 
\bibitem[\protect\citeauthoryear{Teplitz et al.}{2005}]{Teplitz05} Teplitz H.I., Charmandaris V., Chary R., Colbert J.W., Armus L., Weedman D., 2005, ApJ, 634, 128 
\bibitem[Teplitz et al.(2007)]{2007ApJ...659..941T} Teplitz, H.~I., et al.\ 2007, \apj, 659, 941 
\bibitem[Tran et al.(2001)]{2001ApJ...552..527T} Tran, Q.~D., et al.\ 2001, \apj, 552, 527 
\bibitem[Valiante et al.(2007)]{2007ApJ...660.1060V} Valiante, E., Lutz, D., Sturm, E., Genzel, R., Tacconi, L.~J., Lehnert, M.~D., \& Baker, A.~J.\ 2007, \apj, 660, 1060
\bibitem[\protect\citeauthoryear{Yan et al.}{2005}]{2005ApJ...628..604Y} Yan L., et al., 2005, ApJ, 628, 604 
\bibitem[Yan et al.(2007)]{2007ApJ...658..778Y} Yan, L., et al.\ 2007, \apj, 658, 778 


\end{thebibliography}
\end{document}